\documentclass{aa}  

\usepackage{graphicx}
\usepackage{txfonts}
\usepackage{multirow}
\usepackage{placeins}
\begin{document}

   \title{Ice as a Photochemical Shield: Adsorption Energetics and
          Spectroscopic Modulation of Interstellar Thiocyanates
          \object{HCSCN} and \object{HCSCCH} in \object{TMC-1}}

   \subtitle{}

   \author{S. G. Dastider\inst{1} (saptarshhigdask@gmail.com)
        \and A. S. Negi\inst{2}
        \and K. Mondal\inst{2}\email
        \and J. Cyriac\inst{1}
        }

   \institute{
      Department of Chemistry, Indian Institute of Space Science and
      Technology, India
      \and
      Department of Physics and Astrophysics, University of Delhi, India
   }

   \date{Received \today}

 
  \abstract
   {The recent detections of thioformyl cyanide (HCSCN) and propynethial (HCSCCH) in the Taurus Molecular Cloud-1 (TMC-1) provide critical insights into the interstellar sulfur inventory. However, the sequestration mechanisms and survivability of these complex S-bearing organics on dust grain mantles remain poorly constrained, limiting our understanding of the missing sulfur problem. In this work, we present a comprehensive computational study characterizing the site-specific adsorption of HCSCN and HCSCCH on amorphous solid water (ASW), modeled via (H$_2$O)$_{n=6-16}$ clusters. Ground-state geometries, binding energies ($E_{\text{des}}$), and vibrational Stark shifts were calculated at the $\omega$B97XD/def2-TZVP level of theory, corroborated by Quantum Theory of Atoms in Molecules (QTAIM) topological analyses. Vertical electronic excitations were further evaluated via TD-DFT to assess solvatochromic perturbations. Our results reveal a highly heterogeneous binding environment, with desorption energies spanning a broad distribution ($\approx 1500$ K to $4900$ K). Strong cavity binding sites, dominated by cooperative hydrogen-bonding networks, induce significant Stark shifts in the $C=S$ stretching modes. Crucially, while the ice matrix exerts a negligible solvatochromic shift on the UV transition wavelengths, strongly bound configurations exhibit a pronounced hyperchromic enhancement of the oscillator strength. Implementing these site-specific thermodynamic parameters into the UCLCHEM gas-grain code demonstrates that these species do not undergo a singular sublimation event. Instead, the heterogeneous ice topology dictates a gradual thermal desorption profile during hot-core warm-up phases. Furthermore, the hyperchromic effect establishes a survival paradox: while deeply trapped populations are thermodynamically shielded against thermal desorption, they possess larger UV absorption cross-sections, rendering them highly vulnerable to photodissociation by the interstellar radiation field prior to sublimation.}

   \authorrunning{S. G. Dastider}
   \titlerunning{Ice as a Photochemical Shield}

   \maketitle
%

\section{Introduction}

The detection of complex molecular species in the interstellar medium (ISM) continues to fundamentally challenge and refine established models of chemical evolution in cold, dense molecular clouds \cite{annurev_astro, iff_publications}. A central, unresolved paradigm in contemporary astrochemistry is the ``missing sulfur problem''---the persistent observation that gas-phase sulfur abundances in dense clouds are depleted by orders of magnitude relative to their established cosmic abundances in the diffuse ISM \cite{cernicharo2021_sulfursaga}. This severe depletion strongly suggests that the bulk of the active sulfur inventory is sequestered as refractory material or complex organics on the icy mantles of interstellar dust grains, rather than remaining in the gas phase \cite{hs2_detection, crab_nebula_arh}.

Among the most significant recent observational breakthroughs are the discoveries of thioformyl cyanide (HCSCN) and propynethial (HCSCCH) in the Taurus Molecular Cloud-1 (TMC-1) \cite{cernicharo2021_sulfursaga}. The detection of these specific carbon-chain sulfur organics in TMC-1, the region increasingly recognized for its profound chemical complexity \cite{indene_tmc1} alongside related species such as thioacetaldehyde \cite{agundez2025_thioacetaldehyde}, provides a unique window into the competitive differentiation between sulfur and oxygen during the early stages of star formation. Furthermore, driven by rapidly expanding submillimeter wave spectroscopy catalogs \cite{submm_spectroscopy}, the continued identification of novel S-bearing molecules, including HC(S)NC \cite{hcsnc_detection} and \textit{trans}-HC(O)SH \cite{hcosh_formation}, highlights the sheer density of the sulfur reaction network. While computational studies have explored gas-phase formation routes, such as the reactions of H$_2$CS and H$_2$CO with the CN radical \cite{gas_phase_h2cs_cn_pmc, gas_phase_h2cs_cn_rg}, gas-phase pathways alone are insufficient to explain the observed abundances, necessitating a rigorous treatment of grain-surface chemistry.

Interpreting these observed abundances requires a precise understanding of the mechanisms governing molecular partitioning between the gas phase and the grain surface. Currently, large-scale gas-grain astrochemical kinetic packages---such as UCLCHEM \cite{holdship2017_uclchem, uclchem_gasgrain}, NEATH \cite{neath_code}, and Chempl \cite{chempl_code}---are employed alongside comprehensive reaction databases like the GRETOBAPE network \cite{gretobape_network} to simulate cloud evolution. However, to calculate thermal evaporation rates during the warm-up phase of a protostellar core, these macroscopic models frequently approximate interstellar dust grains as uniform, homogenous spheres, relying on a single, static binding energy for each adsorbate species. 

In reality, interstellar ice is primarily composed of amorphous solid water (ASW), which is inherently disordered, highly heterogeneous, and structurally complex \cite{water_clusters_taylor, clusters_classical_water}. Finite water clusters serve as highly effective, computationally tractable proxies for exploring this surface complexity \cite{accurate_water_clusters, md_hydrate}. Extensive theoretical and experimental determinations of structural motifs, ranging from hexagonal networks to undecamer clusters, demonstrate that the ice surface presents a multitude of binding configurations \cite{hexagonal_water, undecamer_water}. Consequently, molecular adsorption is fundamentally site-specific; the binding energy of an adsorbate varies drastically depending on its specific local coordination environment \cite{binding_sn2024}. Relying on a single averaged binding energy severely misrepresents the thermodynamic realities of molecular retention near protostellar snowlines \cite{bulik2025_d5cp, benzene_binding_ice, cs2_hydrogenation_ice}.

To accurately capture the physics of these non-covalent interactions, researchers utilize classical empirical potentials such as the TIP4P and the explicitly tetrahedral TIP5P rigid water models \cite{wales2005_tip5p, tip5p_reopt}. In this work, we present a comprehensive computational framework to investigate the site-specific adsorption of HCSCN and HCSCCH on ASW, utilizing (H$_2$O)$_{n=6-16}$ cluster models. Ground-state binding energies are derived at a high level of theory using dispersion-corrected Density Functional Theory (DFT), yielding highly accurate vibrational spectroscopic features and adsorption energies that are benchmarked against current literature standards \cite{bulik2025_d5cp, perrero2024_d4cp}. Crucially, we couple this thermodynamic data with Time-Dependent DFT derived UV-Vis absorption cross-sections to systematically assess solvatochromic perturbations induced by the ice matrix.

By integrating these site-specific binding and sublimation parameters into the UCLCHEM code, we enable a significantly more realistic simulation of the sulfur chemistry network. This unified approach reveals a fundamental ``survival paradox'' governing sulfur organics. We demonstrate that the heterogeneous topology of ASW creates a gradual thermal desorption window, while deeply bound cavity sites induce a highly molecule-specific hyperchromic enhancement of UV oscillator strength. Ultimately, coupling the microscopic details of hydrogen bonding with the macroscopic dynamics of molecular clouds brings the field closer to identifying the long-sought carriers of the missing interstellar sulfur.
\section{Computational Methodology}

\subsection{Ice Cluster Modeling and Visualization}
To simulate the structurally heterogeneous environment of amorphous solid water (ASW), finite water clusters (H$_2$O)$_{n}$ with $n = 6, 8, 10, 12,$ and $16$ were employed as local surface proxies. The initial geometries for these clusters were constructed based on the established global minima for the TIP4P \cite{tip4p_original} and explicitly tetrahedral TIP5P \cite{wales2005_tip5p} empirical water potentials. The adsorbates, thioformyl cyanide (HCSCN) and propynethial (HCSCCH), were introduced to various structurally distinct sites on the clusters---ranging from exposed surface terraces to deeply coordinated hydrogen-bonding cavities to ensure a comprehensive sampling of the adsorption phase space. All structural manipulation, initial complex generation, and subsequent visual rendering of the optimized binding motifs were performed using the Atomic Simulation Environment (ASE) \cite{ase_paper}.

\subsection{Ground State Quantum Chemical Calculations}
All quantum chemical calculations were performed using the Gaussian 09 software package \cite{g09}. Ground-state geometry optimizations and harmonic vibrational frequency calculations were executed within the framework of Density Functional Theory (DFT). TAll electronic structure calculations were performed using the $\omega$B97X-D range-separated hybrid functional \cite{wb97xd_paper} in conjunction with the def2-TZVP basis set \cite{def2_basis}. The selection of this specific level of theory is driven by recent comprehensive benchmarking in astrochemical modeling. As demonstrated by \citet{Ferrero_2021} and \citet{Sameera_2021}, the inclusion of empirical dispersion corrections within the $\omega$B97X-D functional is strictly required to accurately capture the shallow potential energy surfaces, long-range van der Waals interactions, and cooperative hydrogen-bonding networks governing ice-adsorbate complexes. This functional, paired with the sufficiently large def2-TZVP basis set, provides an accuracy comparable to computationally expensive Coupled Cluster methods, making it highly robust for evaluating the structural heterogeneity of extensive water-ice clusters without prohibitive computational costs.

Harmonic vibrational frequencies were computed at the same level of theory to confirm that all optimized structures correspond to true energetic minima (characterized by the absence of imaginary frequencies) and to extract the zero-point energy (ZPE) corrections.

The site-specific desorption energy ($E_{\text{des}}$) for each complex was rigorously evaluated to ensure accurate astrochemical kinetics. To account for the artificial overestimation of interaction energies inherent to finite basis sets, the standard Boys-Bernardi Counterpoise (CP) method \cite{bsse_cp} was utilized. The fully corrected binding energy ($E_{\text{binding}}$) was calculated by taking the Counterpoise-corrected complexation energy and subtracting the difference in zero-point energy induced by the adsorption, defined as:
\begin{equation}
    E_{\text{binding}} = E_{\text{complexation}}^{\text{CP}} - \Delta E_{\text{ZPE}}
    \label{eq:binding_energy}
\end{equation}

where the net zero-point energy difference ($\Delta E_{\text{ZPE}}$) is given by:
\begin{equation}
    \Delta E_{\text{ZPE}} = E_{\text{ZPE}}^{\text{complex}} - \left( E_{\text{ZPE}}^{\text{ice}} + E_{\text{ZPE}}^{\text{adsorbate}} \right)
    \label{eq:delta_zpe}
\end{equation}

Here, $E_{\text{ZPE}}^{\text{complex}}$, $E_{\text{ZPE}}^{\text{ice}}$, and $E_{\text{ZPE}}^{\text{adsorbate}}$ represent the unscaled zero-point energies of the optimized interacting complex, the isolated water-ice cluster, and the isolated gas-phase adsorbate, respectively. 

For implementation into the macroscopic kinetic network, the thermal desorption energy ($E_{\text{des}}$) is taken as the absolute magnitude of this thermodynamic barrier:
\begin{equation}
    E_{\text{des}} = |E_{\text{binding}}|
    \label{eq:desorption_energy}
\end{equation}

To quantify the physical perturbations induced by the ice matrix on the trapped species, we evaluated both the infrared (IR) and ultraviolet (UV) spectroscopic deviations relative to the isolated gas-phase molecule. The IR vibrational Stark shift ($\Delta\nu$) for the primary functional groups is defined as:
$$\Delta\nu = \nu_{\text{adsorbed}} - \nu_{\text{gas}}$$
where $\nu$ represents the harmonic vibrational frequency. 

Similarly, the UV-Vis absorption properties, evaluated via Time-Dependent DFT (TD-DFT), were analyzed to ascertain the influence of the ice cavity on the electronic transitions. The shift in the maximum absorption wavelength ($\Delta\lambda$) is given by:
$$\Delta\lambda = \lambda_{\text{adsorbed}} - \lambda_{\text{gas}}$$
Furthermore, to model the enhanced \textit{in situ} photodestruction central to the ``Survival Paradox,'' the hyperchromic effect was quantified by calculating the relative percentage change in the transition oscillator strength ($\Delta f$):
$$\Delta f (\%) = \left( \frac{f_{\text{adsorbed}} - f_{\text{gas}}}{f_{\text{gas}}} \right) \times 100$$
This fractional change in oscillator strength ($\Delta f$) is subsequently mapped directly to the scaling of the photodissociation rate ($\alpha$) within our macroscopic UCLCHEM kinetic models.

\subsection{Topological and Non-Covalent Interaction (NCI) Analysis}

To physically categorize the nature of the specific host-guest interactions (e.g., localized hydrogen bonding versus broad dispersive contacts) and to definitively rule out chemisorption, a dual-topological approach was employed using the Multiwfn wavefunction analyzer program \cite{multiwfn}. 

First, the Quantum Theory of Atoms in Molecules (QTAIM) \cite{qtaim_bader} was applied to evaluate the fundamental topology of the electron density. Key parameters, specifically the electron density ($\rho$) and the Laplacian of the electron density ($\nabla^2\rho$) were extracted at the inter-molecular Bond Critical Points (BCPs) formed between the adsorbate and the surrounding ice matrix.

Furthermore, to visualize the spatial extent and relative strength of these interactions, the Non-Covalent Interaction (NCI) index \cite{johnson_nci_2010} was utilized. NCI analysis maps the Reduced Density Gradient (RDG) to identify regions of low electron density and low density gradient characteristic of non-covalent interactions. The physical nature of the binding domains was characterized by projecting the electron density multiplied by the sign of the second Hessian eigenvalue, $\text{sign}(\lambda_2)\rho$, onto the RDG isosurfaces. This allowed for a clear visual and quantitative distinction between strongly attractive, highly localized hydrogen bonds (e.g., deep cavity stabilization) and weaker, delocalized van der Waals forces (e.g., surface physisorption).

\subsection{Excited State Calculations and UV-Vis Spectroscopy}
To investigate the solvatochromic perturbations and the hyperchromic enhancement induced by the ice matrix, vertical electronic excitations were computed utilizing Time-Dependent Density Functional Theory (TD-DFT). Single-point TD-DFT calculations were performed on the $\omega$B97XD/def2-TZVP optimized ground-state geometries, as electronic transitions occur on a timescale significantly faster than nuclear relaxation (consistent with the Franck-Condon principle). Excitation wavelengths ($\lambda$) and oscillator strengths ($f$) were extracted to construct the theoretical UV-Vis absorption cross-sections. To ensure the robustness of the observed hyperchromic trends and to rule out functional-dependent artifacts regarding charge-transfer excitations, the extreme thermodynamic binding configurations were further benchmarked using the CAM-B3LYP functional \cite{camb3lyp_paper}.

\subsection{Astrochemical Kinetic Modelling}
\label{sec:uclchem_methods}

Gas-grain chemical simulations were performed using the
open-source kinetic code UCLCHEM
\citep{holdship2017_uclchem, uclchem_gasgrain} to evaluate
the macroscopic astrochemical consequences of the computed
site-specific binding energies and TD-DFT oscillator
strength enhancements. The chemical evolution was modelled
via a standard two-phase hot-core approach following the
framework of \citet{Garrod2008}.

\subsubsection*{Phase 1: Prestellar Cloud Collapse}

Phase~1 models the isothermal free-fall collapse of the
prestellar cloud. The gas density $n_\mathrm{H}$ was
evolved from an initial value of $10^2$~cm$^{-3}$ to a
final dense-core value of $10^5$~cm$^{-3}$, maintained
at a constant dust temperature of 10~K over a total
collapse timescale of $5\times10^6$~yr. This final density
is consistent with observational constraints on the
cyanopolyyne peak of TMC-1, where molecular hydrogen
densities of $10^4$--$10^5$~cm$^{-3}$ are well established
\citep{pratap1997_tmc1, nutter2008_scuba, fuente2019_gems}.
Throughout the collapse, gas-phase species continuously
accrete onto dust grains, building the amorphous solid
water ice mantle. The final abundances at the end of
Phase~1 were saved and used as initial conditions for
Phase~2. The foundational chemical network was adopted
from the UMIST Database for Astrochemistry
\citep{mcelroy2013_umist}, augmented to include gas-phase
formation and grain-surface accretion pathways for HCSCN
and HCSCCH.

\subsubsection*{Phase 2: Protostellar Warm-Up}

Phase~2 simulates the protostellar warm-up. The gas
density was held fixed at $n_\mathrm{H} = 10^5$~cm$^{-3}$
while the dust temperature was ramped linearly from 10~K
to 200~K over a warm-up timescale of
$t_\mathrm{warmup} = 2\times10^5$~yr. The gas-phase
fractional abundance $n/n_\mathrm{H}$, grain-surface
population, and bulk ice population were tracked
simultaneously as functions of dust temperature throughout
the ramp. The warm-up timescale governs the rate at which
the temperature progresses and hence the duration of each
desorption window.

\subsubsection*{Heterogeneous Binding Energy Treatment}

\begin{figure*}
\centering

\begin{minipage}[c]{0.54\textwidth}
\centering
\small
\renewcommand{\arraystretch}{1.25}
\begin{tabular}{llcrrr}
\hline
Molecule & Adsorption site & $N$ &
$E_{\rm des}^{\rm min}$ & $\bar{E}_{\rm des}$ &
$E_{\rm des}^{\rm max}$ \\
 & & & (K) & (K) & (K) \\
\hline
\multicolumn{6}{l}{\textit{TIP4P model}} \\
\hline
\multirow{5}{*}{HCSCN}
  & CN site (cavity) & 5 & 3431 & 3679 & 3899 \\
  & CS site (cavity) & 5 & 2913 & 3319 & 3712 \\
  & Sideways         & 1 & 3010 & 3010 & 3010 \\
  & CN sideways      & 1 & 2106 & 2106 & 2106 \\
  & CS sideways      & 1 & 1989 & 1989 & 1989 \\
\hline
\multirow{4}{*}{HCSCCH}
  & CS site (cavity) & 6 & 1830 & 3182 & 3684 \\
  & CC site          & 3 & 2137 & 2187 & 2253 \\
  & Sideways         & 1 & 2228 & 2228 & 2228 \\
  & CC sideways      & 1 & 1539 & 1539 & 1539 \\
\hline
\multicolumn{6}{l}{\textit{TIP5P model}} \\
\hline
\multirow{2}{*}{HCSCN}
  & CN site (cavity) & 8 & 3396 & 4160 & \textbf{4944} \\
  & CS site          & 9 & 3031 & 3622 & 4138         \\
\hline
\multirow{2}{*}{HCSCCH}
  & CS site (cavity) & 5 & 3025 & 3480 & \textbf{4146} \\
  & CC site          & 4 & 2103 & 2769 & 4073          \\
\hline
\end{tabular}
\end{minipage}
\hfill
\begin{minipage}[c]{0.43\textwidth}
\centering
\includegraphics[width=\linewidth]{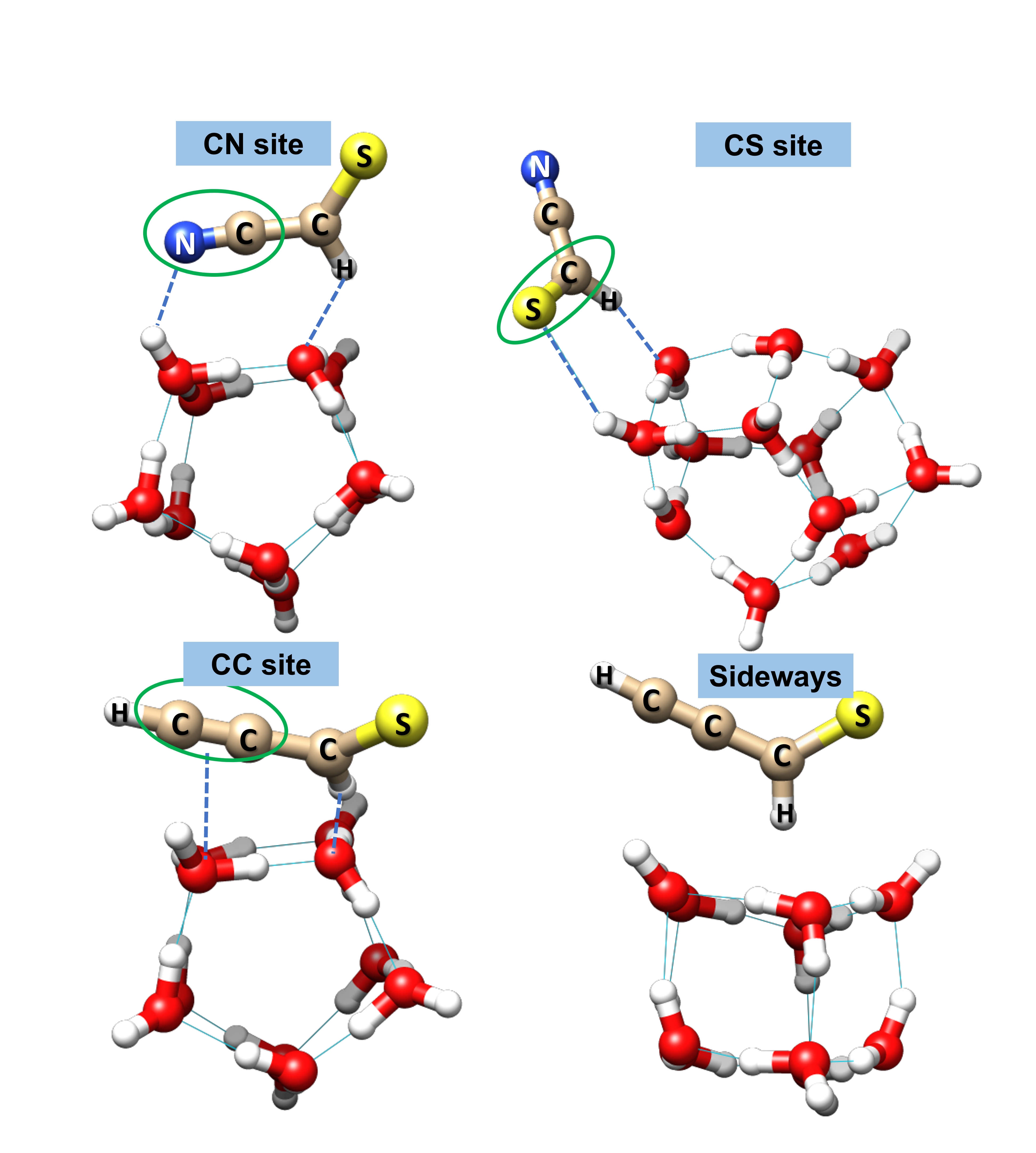}
\end{minipage}

\caption{%
\textit{Left:} \textbf{Table}: Site-specific BSSE-corrected desorption energies
($E_{\rm des}$, in K) for \mbox{HCSCN} and \mbox{HCSCCH} adsorbed on
$(\mathrm{H_2O})_{n=6-16}$ clusters derived from the TIP4P and TIP5P
water potentials at the $\omega$B97X-D/def2-TZVP level.
$N$ denotes the number of optimised configurations per site category.
Boldface values indicate global maxima for each molecule--model
combination.
\textit{Right:} \textbf{Figure} Representative adsorption geometries illustrating the
structurally distinct binding topologies sampled across the
$n = 6$--$16$ survey.
Atom colours: O red, H white, C tan, N blue, S yellow;
cyan distances mark intermolecular hydrogen-bond contacts.}
\label{fig:be_sites}
\end{figure*}

To capture the thermodynamic impact of ASW surface
heterogeneity, the standard single-value binding energy
approximation was replaced by a dual-run approach. For
each species, two independent kinetic evaluations were
performed using the minimum and maximum BSSE-corrected
desorption energies derived from the DFT cluster
calculations (Table of Figure~\ref{fig:be_sites}) as thermodynamic
bounds. Parameterizing a fully continuous binding energy distribution for macroscopic kinetics requires a density of states mapping beyond the scope of finite cluster surveys. Therefore, our dual-run approach is designed specifically to establish the theoretical temporal and thermal boundaries the macroscopic envelope of the sequestration phase. By utilizing the minimum and maximum binding energies, we define the earliest possible thermal release and the maximum possible sequestration timescale. This bounding strategy effectively frames the duration of the 'Waiting Room' phase, during which species are exposed to internal radiation fields.

\subsubsection*{Photodissociation Rate Modification}

To assess the photolytic vulnerability of the
cavity-trapped ice populations, the photodissociation
rate constants within the UCLCHEM network were modified
to reflect the site-dependent UV oscillator strength
perturbations derived from the TD-DFT calculations
(Section~\ref{sec:spectroscopy}). The standard astrochemical
photodissociation rate coefficient takes the form,
\begin{equation}
    k_\mathrm{pd} = \alpha \exp(-\gamma A_V)
    \quad \mathrm{s}^{-1},
    \label{eq:kpd}
\end{equation}
where $\alpha$ is the unattenuated rate in the
standard interstellar radiation field
\citep{draine1978}, $\gamma$ is the dust shielding
parameter, and $A_V$ is the visual extinction.
Since $\alpha \propto \sigma(\nu) \propto f$, where
$\sigma(\nu)$ is the UV absorption cross-section and $f$
is the oscillator strength, the $\alpha$ parameter for
each ice-phase species was scaled by the ratio
$f_\mathrm{adsorbed}/f_\mathrm{gas}$ derived from
the TD-DFT calculations. Independent simulations were
performed for each binding energy extreme, with the
corresponding site-specific oscillator strength ratio
applied consistently. No modification was applied to
gas-phase photodissociation rates, which retain the
unperturbed literature values throughout. It is acknowledged that the structurally disordered ice matrix induces
inhomogeneous line broadening of the absorption bands relative to the
gas phase. However, since the oscillator strength $f$ quantifies the
total integrated transition probability, the linear scaling of
$k_{\rm pd}$ by $f_{\rm adsorbed}/f_{\rm gas}$ captures the net
enhancement in total absorbed photons irrespective of matrix-induced
spectral redistribution, provided the interstellar radiation field
lacks narrow emission features coincident with the absorption band, which is a condition satisfied by the broadband Draine field
\citep{draine1978}.
Furthermore, deep within the dense core during the Phase~2 warm-up,
the external UV field is heavily attenuated by the dust column
($A_V \gtrsim 10$\,mag). The modelled photolysis therefore operates
primarily through the \emph{internal} cosmic-ray-induced secondary UV
field \citep{prasad1983}, which acts on the same scaled cross-sections
and maintains the hyperchromic penalty throughout the Waiting Room
interval.

\begin{table*}
\centering
\caption{TD-DFT ($\omega$B97XD/def2-TZVP) UV oscillator strengths ($f$) and
vibrational Stark shifts ($\Delta\nu$, cm$^{-1}$) for HCSCN and HCSCCH
adsorbed on (H$_2$O)$_{n=6-16}$ clusters. Gas-phase reference oscillator strengths:
$f_\mathrm{gas}(\mathrm{HCSCN}) = 0.2802$; $f_\mathrm{gas}(\mathrm{HCSCCH}) = 0.3029$.
$\Delta f/f_\mathrm{gas}$ (\%) denotes the mean hyperchromic/hypochromic change.
Positive $\Delta\nu_\mathrm{C=S}$ values indicate a blueshift; negative values a redshift.
Bold entries mark, within each molecule–model block, the minimum and maximum values
for each of the three right-most columns.}
\label{tab:uv_ir}
\begin{tabular}{lllcrrrr}
\hline
Molecule & Model & Adsorption site & $N$ & $\bar{f}$ & $\Delta f/f_\mathrm{gas}$ (\%) & $\Delta\nu_\mathrm{C=S}$ (cm$^{-1}$) & $\Delta\nu_\mathrm{C{\equiv}N/C{\equiv}C}$ (cm$^{-1}$) \\
\hline
\multirow{5}{*}{HCSCN} & \multirow{5}{*}{TIP4P}
  & CN site (cavity)  & 5 & 0.307 &   $+11.7$) & $\mathbf{+0.5}$  & $+3.2$ \\
  &                   & CS site (cavity)  & 5 & 0.265 & $-5.4$                        & $\mathbf{-11.8}$ & $\mathbf{+4.2}$ \\
  &                   & Sidewise          & 1 & 0.220 & $\mathbf{-21.3}$              & $-5.8$          & $+2.0$ \\
  &                   & CN sideways       & 1 & 0.274 & $-2.1$                        & $-8.4$          & $\mathbf{+0.4}$ \\
  &                   & CS sideways       & 1 & 0.245 & $-12.6$                       & $-11.5$         & $+2.8$ \\
\hline
\multirow{2}{*}{HCSCN} & \multirow{2}{*}{TIP5P}
  & CN site (cavity)  & 8 & 0.290 & $\mathbf{+3.5}$                & $\mathbf{+5.1}$  & $\mathbf{+4.0}$ \\
  &                   & CS site           & 9 & 0.239 & $\mathbf{-14.7}$              & $\mathbf{-3.6}$ & $\mathbf{+4.5}$ \\
\hline
\multirow{4}{*}{HCSCCH} & \multirow{4}{*}{TIP4P}
  & CS site (cavity)  & 6 & 0.299 & $-1.4$                        & $\mathbf{-5.3}$ & $+3.4$ \\
  &                   & CC site           & 3 & 0.300 & $\mathbf{-0.8}$              & $\mathbf{+0.7}$ & $\mathbf{+10.4}$ \\
  &                   & Sideways          & 1 & 0.270 & $\mathbf{-11.0}$             & $-2.6$         & $+4.4$ \\
  &                   & CC sideways       & 1 & 0.299 & $-1.2$                        & $-4.1$         & $\mathbf{-0.3}$ \\
\hline
\multirow{2}{*}{HCSCCH} & \multirow{2}{*}{TIP5P}
  & CS site (cavity)  & 5 & 0.299 & $\mathbf{-1.4}$               & $\mathbf{-4.8}$ & $\mathbf{+6.4}$ \\
  &                   & CC site           & 4 & 0.301 & $\mathbf{-0.6}$               & $\mathbf{+2.6}$ & $\mathbf{+11.4}$ \\
\hline
\end{tabular}
\end{table*}



\section{Results and Discussion}

\subsection{Binding Energies and Ice Topology}\label{sec:be}

\begin{figure*}
    \centering
    \includegraphics[width=\textwidth]{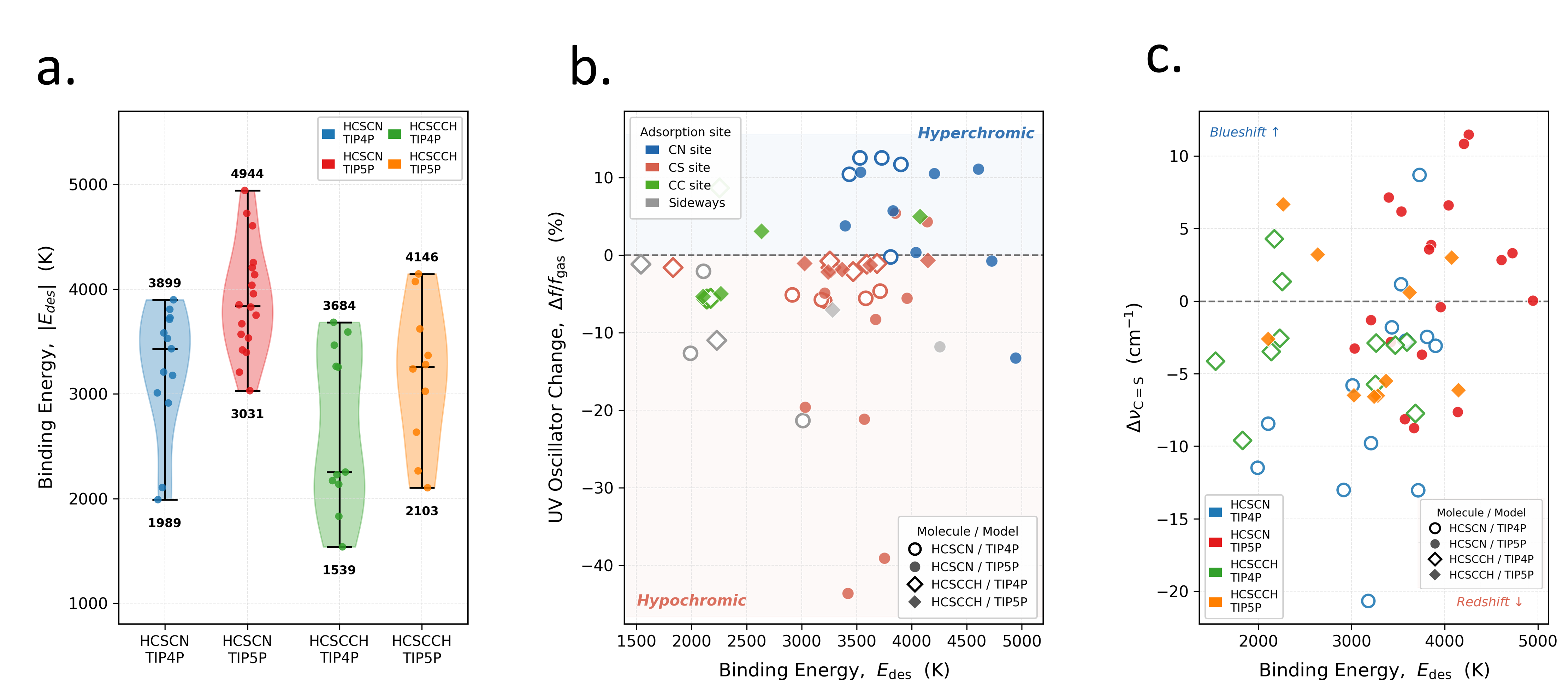}
    \caption{Binding energies, UV oscillator strengths, and C=S IR Stark shifts 
    for HCSCN and HCSCCH adsorbed on amorphous solid water (ASW) ice clusters. 
    a. Violin plots with overlaid strip charts showing the distribution 
    of binding energies (in Kelvin) across all optimized adsorption geometries for 
    HCSCN/TIP4P, HCSCN/TIP5P, HCSCCH/TIP4P, and HCSCCH/TIP5P model ice clusters. 
    b. Scatter plot of binding energy versus percentage change in 
    TD-DFT oscillator strength (\textit{f}) relative to the gas-phase reference, 
    colour-coded by adsorption site (CN-site: cyan; CS-site: gold; sideways: grey). 
    Filled circles denote HCSCN; open squares denote HCSCCH. 
    c. C=S stretching frequency shift ($\Delta\nu$, cm$^{-1}$) as a 
    function of binding energy for all four systems, illustrating the correlation 
    between adsorption strength and vibrational Stark tuning.}
    \label{fig:main}
\end{figure*}

The central premise of this work is that the intrinsic structural
heterogeneity of amorphous solid water (ASW) renders a single,
averaged binding energy fundamentally insufficient to describe the
thermodynamics of molecular retention on interstellar grain mantles.
By sampling the full topological phase space of
$(H_2O)_{n=6-16}$ clusters from exposed surface terraces to
deeply coordinated hydrogen-bonding cavities, we recover a broad,
continuous distribution of desorption energies for both target
molecules (Fig.~\ref{fig:main}a; Table~\ref{fig:be_sites}).

\subsubsection*{Thermodynamic landscape and bonding topology on
TIP4P clusters}

Initial geometry optimizations and binding energy evaluations on
TIP4P-derived clusters reveal a remarkably broad thermodynamic
distribution, governed entirely by the local coordination
environment of the adsorbate rather than cluster size alone.
For HCSCN, the BSSE-corrected desorption energies span nearly
2000~K on a single cluster: from $E_\mathrm{des} \approx 1989$~K
for weakly bound, sulfur-anchored CS-sideways orientations to
$E_\mathrm{des} = 3899$~K for the deepest CN-coordinated cavity
site on the 10-molecule cluster (Table~\ref{fig:be_sites}).
HCSCCH exhibits an equally broad distribution, ranging from
$E_\mathrm{des} = 1539$~K for CC-sideways geometries to
$E_\mathrm{des} = 3684$~K for CS-cavity coordination
(Table~\ref{fig:be_sites}).
The physical origin of this spread is made explicit in
Fig.~\ref{fig:tip4p10_sites}, which presents the three
representative binding topologies --- CN-cavity, CS-cavity,
and sideways surface terrace on the TIP4P $(H_2O)_{10}$
cluster for both molecules.
For HCSCN, embedding the nitrile terminus within a multi-centre
donor cavity (panel~a; $E_\mathrm{des} = 3899.1$~K) yields
$\sim$190~K additional stabilisation over the CS-anchored geometry
(panel~b; $E_\mathrm{des} = 3712.0$~K) and $\sim$890~K over the
weakly coordinated sideways orientation (panel~c;
$E_\mathrm{des} = 3010$~K).
For HCSCCH, the CS-cavity site (panel~e;
$E_\mathrm{des} = 3684.2$~K) exceeds the CC-site (panel~d;
$E_\mathrm{des} = 2137$~K) and sideways geometry (panel~f;
$E_\mathrm{des} = 2228$~K) by up to $\sim$1550~K.

To understand the microscopic origin of these site-dependent
energetic and spectroscopic differences, we characterise the
nature of the adsorbate-ice interactions using QTAIM topological
analysis and Non-Covalent Interaction (NCI) index mapping.
For clarity and interpretability, we demonstrate the full
topological analysis on the TIP4P $(H_2O)_6$ system which the
smallest cluster that reproduces the complete CN-site and CS-site
bonding motifs; before establishing that the same structural
hierarchy persists quantitatively across all cluster sizes
$n = 6$--$16$ (Table~\ref{fig:be_sites}).

\begin{figure*}
    \centering \includegraphics[width=0.8\textwidth]{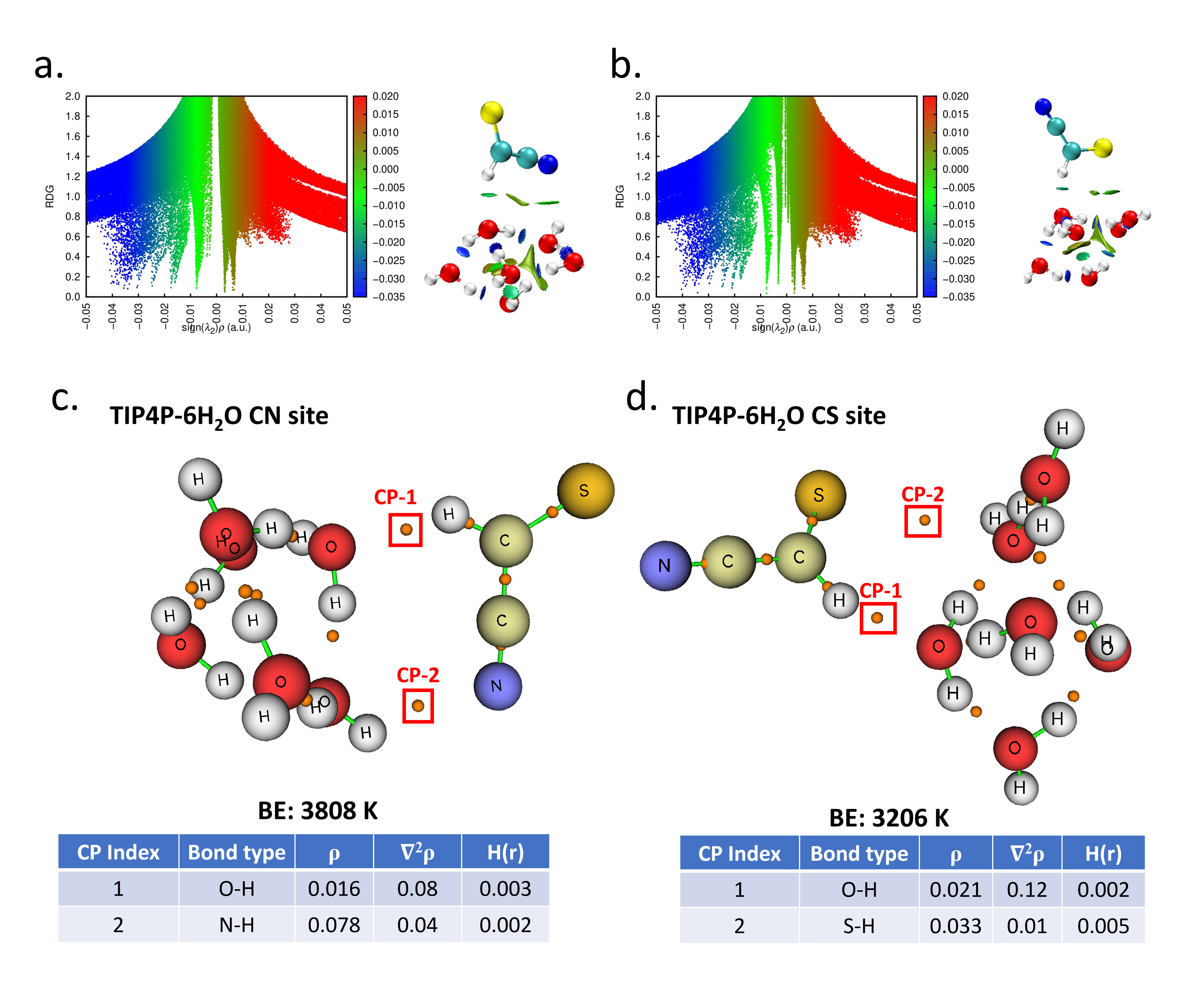}
    \caption{NCI and QTAIM bond topology analysis for \mbox{HCSCN} on the
TIP4P $(\mathrm{H_2O})_6$ cluster at the CN site ($E_\mathrm{des} = 3808$\,K,
panels~a,c) and CS site ($E_\mathrm{des} = 3206$\,K, panels~b,d).
Top: RDG scatter plots with isosurface insets ($s = 0.5$\,a.u.).
Bottom: QTAIM molecular graphs; intermolecular BCPs are labelled CP-1 and
CP-2 with $\rho$, $\nabla^2\rho$, and $H(r)$ values tabulated below
each panel}
    \label{fig:qtaim}
\end{figure*}

The TIP4P $(H_2O)_6$ system is chosen as the reference
system for the topological analysis because it represents
the smallest cluster that simultaneously sustains both a
CN-coordinated and a CS-coordinated cavity motif with fully resolved. Critically, the energetic gap recovered from CN-CS in the
TIP4P $(H_2O)_6$ system ($\Delta E_\mathrm{des} \approx 602$~K;
Table~\ref{fig:be_sites}) is quantitatively representative of
the gap in the full survey $n = 6$ -- - $16$ 
($\Delta\bar{E}_\mathrm{des} \approx 580$ -- - $630$~K),
confirming that the electronic structure of the binding
interface converges at this cluster size even as the
absolute desorption energies continue to grow with $n$.
The limited system size also renders the QTAIM molecular
graph and NCI isosurface directly interpretable: with only
six donor molecules contributing, every BCP and every RDG
island in Fig.~\ref{fig:qtaim} can be unambiguously
assigned to a specific intermolecular contact, providing a
chemically transparent window into the bonding physics that
governs the full binding energy distribution.

The CN-site topology of the TIP4P $(\mathrm{H_2O})_6$ complex is shown in
Fig.~\ref{fig:qtaim}a,c.
When the $-\mathrm{C{\equiv}N}$ terminus is embedded within the donor network
of the ice cavity, QTAIM analysis reveals two intermolecular Bond Critical
Points (BCPs).
The primary contact, CP-1, is an \mbox{O--H$\cdots$N} relay path with
$\rho = 0.016$\,a.u.\ and $\nabla^2\rho = +0.08$\,a.u., indicating a
moderately strong, closed-shell interaction.
The secondary contact, CP-2, is a directional \mbox{N--H$\cdots$N} hydrogen
bond with $\rho = 0.078$\,a.u.\ and $\nabla^2\rho = +0.04$\,a.u., exploiting
the lone pair of the nitrile nitrogen as a potent multi-centre acceptor.
Despite the substantially larger electron density at CP-2, the positive
Laplacian and the small but positive total energy density
$H(r) = +0.002$\,a.u.\ at both BCPs strongly indicate exclusively closed-shell, non-covalent character throughout.
The RDG isosurface (panel~a) confirms this picture: a prominent disc-shaped
green NCI island at negative $\mathrm{sign}(\lambda_2)\rho$ straddling the
\mbox{N$\cdots$H--O} contact visualises the stabilising non-covalent
interaction, flanked by the broader red van-der-Waals envelope of the
enclosing cluster.

At the CS site (Fig.~\ref{fig:qtaim}b,d), the molecule binds via its
thiocarbonyl sulfur, and the QTAIM graph again discloses two BCPs.
CP-1 retains an \mbox{O--H$\cdots$O} character
($\rho = 0.021$\,a.u., $\nabla^2\rho = +0.12$\,a.u., $H(r) = +0.002$\,a.u.),
while CP-2 is an \mbox{S--H$\cdots$S} contact
($\rho = 0.033$\,a.u., $\nabla^2\rho = +0.01$\,a.u., $H(r) = +0.005$\,a.u.).
Despite the larger electron density at CP-2 relative to CP-1, the near-zero
Laplacian at CP-2 signals an interaction approaching the boundary between
closed-shell and shared character --- a direct consequence of the diffuse,
polarisable sulfur lone pairs forming a softer, less directional bond with
the adjacent O--H donor.
Crucially, the absence of a nitrile acceptor means that the CS site cannot
recruit a second strong directional contact equivalent to the
\mbox{N--H$\cdots$N} bridge that distinguishes the CN-site complex; this
structural deficit accounts for the 602\,K deficit in $E_\mathrm{des}$.

Across both sites and all cluster sizes, the Laplacian remains
$\nabla^2\rho \approx +0.01$--$+0.12$\,a.u.\ and the total energy density
$H(r) > 0$\,a.u.\ at all BCPs, confirming that even the deepest cavity
interactions are unambiguously closed-shell and non-covalent throughout.
The extraordinary binding enthalpies of the CN-cavity motifs therefore arise
not from any incipient chemisorption but from the cooperative, synergistic
accumulation of two directional hydrogen-bond contacts that simultaneously
enclose and orient the adsorbate --- a multi-dentate recognition geometry
that the less directional thiocarbonyl sulfur of the CS site cannot replicate.

The extraordinary binding enthalpies of the CN-cavity motifs
therefore arise not from any incipient chemisorption but from the
cooperative, synergistic accumulation of at least two directional
hydrogen-bond contacts that simultaneously enclose and orient the adsorbate.
The stronger CN-site binding reflects the unique dual role of the nitrile
group: it acts as a $\sigma$-acceptor through the nitrogen lone pair
\emph{and} as a $\pi$-acceptor through the $\mathrm{C{\equiv}N}$ bond axis,
enabling the ice cavity to form a multi-dentate, geometrically matched
complement around the polar terminus that the less directional thiocarbonyl
sulfur cannot replicate.
HCSCCH, which entirely lacks the nitrile chromophore, reaches its binding
maximum through CS-coordinated cavity motifs ($E_\mathrm{des} = 3684$~K),
while CC-sideways orientations fall as low as 1539~K.

The CN-cavity $>$ CS-cavity $>$ surface-terrace hierarchy
is reproduced without exception across the entire
$n = 6$--$16$ TIP4P survey (Table~\ref{fig:be_sites}),
confirming that site type rather than cluster size is
the primary determinant of $E_\mathrm{des}$.
For HCSCN, CN-coordinated sites consistently deliver
the highest desorption energies (3431--3899~K), CS-cavity
sites fall 200--600~K below on the same cluster
(2913--3712~K), and sideways orientations drop to
1989--3010~K.
HCSCCH mirrors this pattern with its CS-cavity hierarchy:
CS-coordinated geometries span 3256--3684~K while
CC-sideways configurations fall to 1539--2253~K.
The absolute values fluctuate non-monotonically with $n$
by $\sim$200--400~K within each site class, reflecting
the sensitivity of the local donor-network geometry to
the specific global-minimum topology of each TIP4P
cluster --- a behaviour well-documented in cluster-based
binding energy surveys~\citep{ferrero2020, bulik2025_d5cp}.
What remains invariant is the relative ordering: no
CS-coordinated HCSCN site ever exceeds a CN-coordinated
counterpart on the same cluster, establishing the
site-type hierarchy as a robust, size-independent
property of the adsorbate--ASW interaction.

\subsubsection*{Validation via TIP5P clusters}

Cross-validation on TIP5P clusters reinforces and extends
this picture.
The explicit representation of both $sp^3$ lone pairs on
each water oxygen creates a richer set of geometrically
distinct donor sites --- 4-centre, 5-centre, 4/4-bridge,
and 4/5-bridge cavities --- that are inaccessible to the
planar effective lone-pair approximation of TIP4P,
resulting in a systematically elevated and broader
$E_\mathrm{des}$ distribution (Table~\ref{fig:be_sites}).
For HCSCN, CN-coordinated sites span 3396--4944~K across
$n = 6$--$16$, with the deepest trap being a 5-centre
CN-cavity on the 12-molecule cluster
($E_\mathrm{des} = 4944$~K; Fig.~\ref{fig:tip5p12_sites},
panel~a); CS-coordinated geometries occupy 3031--4138~K,
and bridge motifs unique to TIP5P
(Figs.~\ref{fig:tip5p12_sites}c,~e) sample an
intermediate coordination space absent from the TIP4P
survey.
For HCSCCH, CS-cavity sites span 3025--4146~K while
CC-coordinated geometries fall to 2103--4073~K.
Despite this richer topological diversity, the
CN-cavity $>$ CS-cavity $>$ surface-terrace hierarchy
is preserved without exception across both molecules
and all cluster sizes, and both species remain bracketed
between the CO and NH$_3$ sublimation windows
($\sim$800~K and $\sim$5000~K respectively), consistent
with co-desorption alongside the water-ice mantle at
$\sim$100--130~K during protostellar warm-up.

For HCSCCH, CS-cavity sites span 3025--4146~K while
CC-coordinated geometries fall to 2103--4073~K, with the
deepest trap being a CC-cavity site on the 12-molecule
cluster ($E_\mathrm{des} = 4073$~K).
Despite this richer topological diversity, the
site-type hierarchy established on TIP4P clusters is
preserved without exception: CN-cavity $>$ CS-cavity
$>$ surface terrace for HCSCN, and CS-cavity $\gtrsim$
CC-cavity $>$ sideways for HCSCCH.
Both molecules are bracketed between the sublimation
windows of CO ($E_\mathrm{des} \sim 800$~K) and
NH$_3$ ($E_\mathrm{des} \sim 5000$~K), consistent with
co-desorption alongside the water-ice mantle at
$\sim$100--130~K during protostellar warm-up.
The model-independent preservation of this hierarchy
across both water potentials confirms that the
$\gtrsim$2000~K spread in $E_\mathrm{des}$ per molecule
is a genuine consequence of ASW surface heterogeneity
rather than an artefact of the chosen empirical
potential.

\subsection{Spectroscopic Signatures: IR Stark Shifts and UV Hyperchromism}
\label{sec:spectroscopy}

The heterogeneous binding landscape revealed above is accompanied by equally
heterogeneous perturbations of the molecular electronic and vibrational structure.
These spectroscopic variations---IR Stark shifts and UV solvatochromic/hyperchromic
effects---both serve as observational diagnostics of the adsorption environment
and directly control the photolytic lifetime of the sequestered species.
The full numerical dataset is given in Table~\ref{tab:uv_ir}; the key
correlations are illustrated in Fig.~\ref{fig:main}b--c.

\subsubsection*{Vibrational Stark shifts ($\Delta\nu$)}

Relative Stark shifts ($\Delta\nu = \nu_\mathrm{adsorbed} - \nu_\mathrm{gas}$)
provide site-specific fingerprints that are immune to the systematic harmonic
overestimation of DFT absolute frequencies.  As shown in
Fig.~\ref{fig:main}c and Table~\ref{tab:uv_ir}, the C$=$S stretching mode
carries the largest and most diagnostically informative shifts.  For HCSCN,
CN-cavity sites induce mild blueshifts
($\Delta\nu_\mathrm{C=S} \approx +2$ to $+12$~cm$^{-1}$) because the CN end
bears the mechanical perturbation, relieving strain on the C$=$S bond.
Conversely, CS-cavity and CS-sideways geometries---where the sulfur atom is
directly engaged in hydrogen-bond donation---impose redshifts reaching
$\Delta\nu_\mathrm{C=S} \approx -21$~cm$^{-1}$ for the largest clusters.
The $\mathrm{C{\equiv}N}$ stretch of HCSCN shows universally positive shifts
($+2$ to $+16$~cm$^{-1}$) irrespective of binding site, reflecting the stiffening
of the triple bond upon cage enclosure.

HCSCCH exhibits qualitatively analogous but quantitatively distinct behaviour:
CS-cavity binding produces moderate C$=$S redshifts
($\Delta\nu_\mathrm{C=S} \approx -3$ to $-7$~cm$^{-1}$), while the C$\equiv$C
alkyne stretch is uniformly blueshifted ($+3$ to $+15$~cm$^{-1}$).
The magnitude of the C$=$S redshift is thus a quantitative reporter of direct
sulfur--ice hydrogen-bond engagement, offering a concrete IR marker to
discriminate between CN- and CS-dominated binding topologies in future
JWST ice-mantle absorption spectra.

\subsubsection*{UV solvatochromism and the site-selective hyperchromic effect}

Across all configurations for both molecules, the ASW matrix induces a
negligible solvatochromic shift in the vertical excitation wavelength
($\Delta\lambda < 5$~nm), a finding that is robust across both the
$\omega$B97XD and CAM-B3LYP functionals.  The primary electronic energy gap
is therefore insensitive to the local ice environment.  In stark contrast,
the UV oscillator strength $f$ exhibits a pronounced, \emph{site-selective}
sensitivity (Fig.~\ref{fig:main}b; Table~\ref{tab:uv_ir}).

For HCSCN, CN-cavity sites carry the largest
binding energies and induce a systematic \emph{hyperchromic} enhancement
of the oscillator strength: the mean enhancement across CN-cavity TIP4P
configurations is $\langle\Delta f/f_\mathrm{gas}\rangle \approx +9.4\%$,
while CN-cavity TIP5P configurations yield a more moderate
$\langle\Delta f/f_\mathrm{gas}\rangle \approx +3.5\%$. Individual
CN-cavity structures reach up to $+12.5\%$ (global maximum for TIP4P)
and $+11.7\%$ for the TIP4P$_{10}$ cavity configuration adopted for
detailed analysis, with the most enhanced TIP5P CN-cavity site reaching
$\approx +11.1\%$. By contrast, CS-cavity sites, which also bind
strongly, produce a consistently \emph{hypochromic} response, with mean
enhancements of $\approx -5$ to $-6\%$ for TIP4P and $\approx -15\%$ for
TIP5P. Sideways and other weakly bound configurations are likewise
hypochromic, spanning $\Delta f/f_\mathrm{gas} \approx -2$ to $-21\%$.

This geometry-specific polarisation arises because CN-cavity enclosure
orients the dominant transition dipole moment \emph{parallel} to the
strongest component of the local electric field generated by the
surrounding donor hydrogens, maximising the spatial overlap of the
initial and final state molecular orbitals. The CS-cavity geometry
rotates the chromophore into a more nearly perpendicular orientation,
suppressing the transition moment.

For HCSCCH, this hyperchromic channel is effectively closed. Despite
occupying CS-cavity and CC-site environments of comparable thermodynamic
depth ($E_\mathrm{des} \approx 3200$--4150~K), all HCSCCH configurations
exhibit a predominantly \emph{hypochromic} response, with mean
$\Delta f/f_\mathrm{gas}$ typically within $\approx 0$ to $-2\%$ and the
most extreme sideways configuration reaching $\approx -11\%$. This is
consistent with the absence of a strongly directional nitrile chromophore
and the more delocalised character of the HCSCCH $\pi$ system.

This dichotomy of CN-cavity trapping  increasing the UV cross-section
of HCSCN while leaving HCSCCH unaffected; is the microscopic origin of the
macroscopic Survival Paradox discussed in Section~\ref{sec:astrochem}.
The cluster-size dependence of these spectroscopic
perturbations is summarised in
Figs.~\ref{fig:ir_heatmaps} and~\ref{fig:uv_heatmap_full}
(Appendix~\ref{app:heatmaps}).
The IR heatmaps (Fig.~\ref{fig:ir_heatmaps}) demonstrate
that the site-specific Stark shift patterns identified
above are persistent across all cluster sizes from
$n = 6$ to $n = 16$, ruling out a size-dependent
artefact in the vibrational response.
The UV heatmap (Fig.~\ref{fig:uv_heatmap_full}, right
panel) confirms that the net hyperchromic signal for
HCSCN is cluster-size-consistent at CN-dominated
configurations.
The two strongly negative cells are the HCSCN/TIP5P $n=6$
site 3 ($\Delta f = -43.6\%$) and HCSCN/TIP5P $n=12$
site 2 ($\Delta f = -39.0\%$) which are not indicative of
genuine UV quenching but instead arise from a
\textit{Davydov-type excited-state splitting} the detailed have been discussed in the next section and 
these entries are therefore excluded from the
site-averaged statistics and treated individually in
Table~\ref{tab:hcscn_tip5p} (see footnote $^\dagger$);
their anomalous character is confirmed as
geometry-specific rather than a systematic cluster-size
effect, since no analogous splitting is observed at
any other TIP5P configuration across $n = 6$--$16$.
The full cluster-size-resolved Stark shift distributions
are presented in Appendix~\ref{app:heatmaps},
Fig.~\ref{fig:ir_heatmaps}; the site-averaged values
used in the main analysis are given in
Tables~\ref{tab:hcscn_tip4p}--\ref{tab:hcscch_tip5p}.

\subsubsection*{Natural Transition Orbital Analysis and the Origin of Site-Selective Oscillator Strength Modulation}

\begin{figure}
    \centering
    \includegraphics[width=\linewidth]{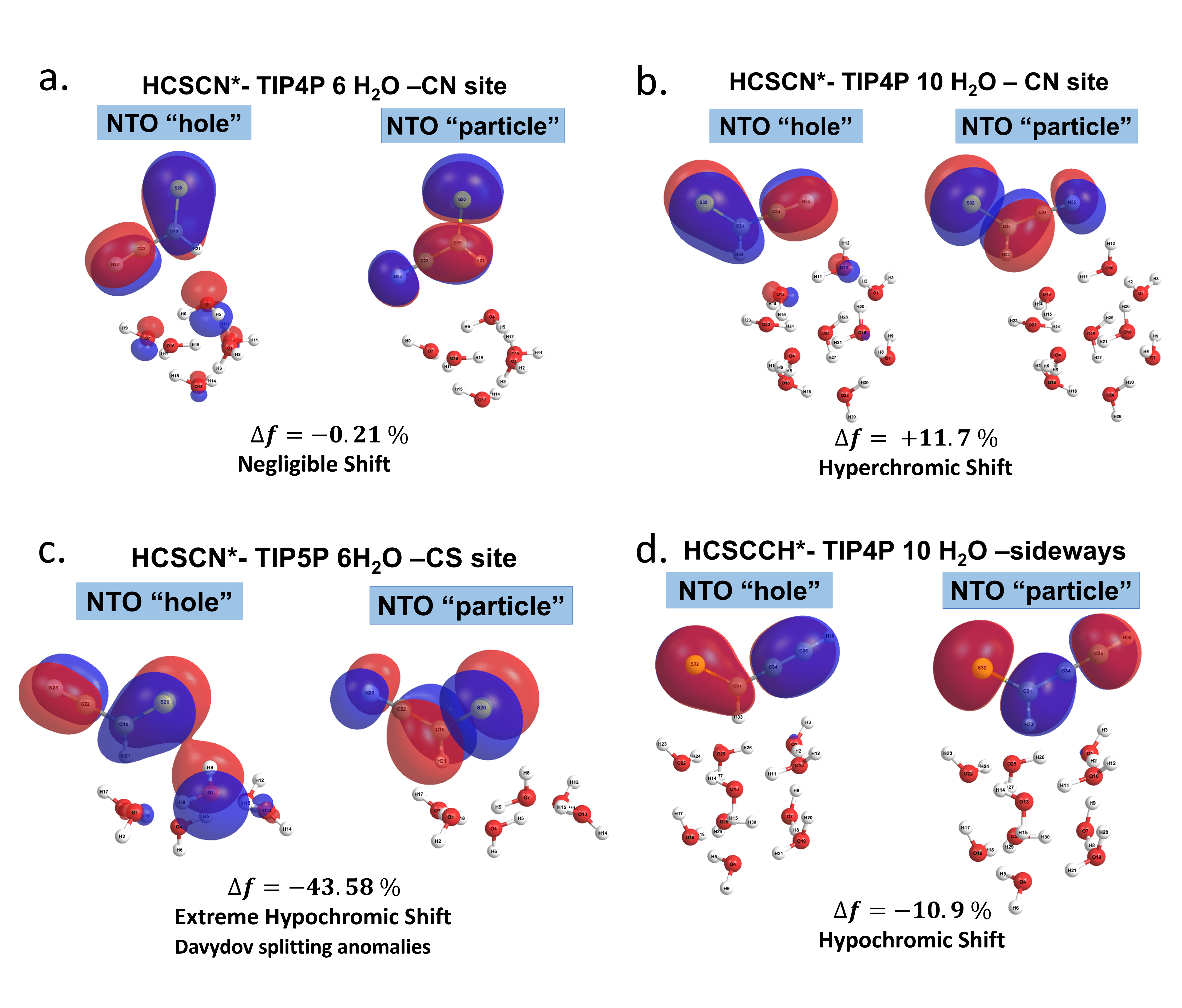}
    \caption{Natural Transition Orbital (NTO) hole--particle pairs for the
  four representative adsorption configurations.
  (a)~\mbox{HCSCN}--TIP4P $6\,\mathrm{H_2O}$, CN site,
      $\Delta f = -0.21\%$ (negligible shift).
  (b)~\mbox{HCSCN}--TIP4P $10\,\mathrm{H_2O}$, CN site,
      $\Delta f = +11.7\%$ (hyperchromic enhancement).
  (c)~\mbox{HCSCN}--TIP5P $6\,\mathrm{H_2O}$, CS site,
      $\Delta f = -43.58\%$ (Davydov splitting anomaly).
  (d)~\mbox{HCSCCH}--TIP4P $10\,\mathrm{H_2O}$, sideways,
      $\Delta f = -10.9\%$ (moderate hypochromism).
  In each panel the hole and particle orbitals correspond to the dominant
  $\mathrm{S_0 \rightarrow S_1}$ transition (largest NTO singular value).
  Isosurface value $= 0.02\,\mathrm{a.u.}$;}
    \label{fig:nto}
\end{figure}

To elucidate the electronic structural origin of the site-selective hyperchromic and hypochromic responses described above, Natural Transition Orbital \citep[NTO;][]{martin2003nto} decompositions were computed for four representative configurations spanning the full range of observed photophysical behaviour: (i) a negligible-shift case (\mbox{HCSCN}--TIP4P $6\,\mathrm{H_2O}$, CN site, $\Delta f = -0.21\%$), (ii) the canonical hyperchromic case (\mbox{HCSCN}--TIP4P $10\,\mathrm{H_2O}$, CN site, $\Delta f = +11.7\%$), (iii) the extreme hypochromic/Davydov-split anomaly (\mbox{HCSCN}--TIP5P $6\,\mathrm{H_2O}$, CS site, $\Delta f = -43.58\%$), and (iv) a moderate hypochromic case (\mbox{HCSCCH}--TIP4P $10\,\mathrm{H_2O}$, sideways, $\Delta f = -10.9\%$). In each panel of Fig.~\ref{fig:nto}, the NTO hole and particle orbitals depicted correspond to the dominant $\mathrm{S_0 \rightarrow S_1}$ transition, i.e.\ the pair carrying the largest singular value (highest oscillator strength contribution) within the NTO expansion of that specific complex. 

In each panel the NTO pair depicted corresponds to the dominant
$\mathrm{S_0 \rightarrow S_1}$ transition, i.e.\ the pair carrying the
largest singular value within the NTO expansion of that complex.
The key result is unambiguous: only CN-cavity adsorption of \mbox{HCSCN}
generates a cavity-aligned reorganisation of the transition dipole moment
$\boldsymbol{\mu}_{ge}$, in which the particle orbital acquires amplitude
on the nitrogen lone-pair axis and the proximal O--H donors of the
enclosing water cage, amplifying the spatial overlap integral
$\langle\psi_g|\hat{r}|\psi_e\rangle$ and hence $f$ and $k_\mathrm{pd}$.
All \mbox{HCSCCH} configurations and CS-cavity sites for both molecules
either leave $f$ essentially unchanged or suppress it moderately, because
the alkyne chromophore and the diffuse thiocarbonyl sulfur cannot sustain
the multi-centre cavity enclosure that drives this alignment.
The two strongly negative outliers  TIP5P $n=6$ site~3 and
TIP5P $n=12$ site~2 ($\Delta f = -43.6$ and $-39.0\%$ respectively)
do not represent genuine UV quenching.
Rather, the tightly directional sp$^3$ lone pairs of the TIP5P donor
network at these specific CS-site geometries place two O--H oscillators
in near-resonant electrostatic coupling with the primary
$\pi\rightarrow\pi^*$ transition of \mbox{HCSCN}, inducing a
Davydov-type excited-state splitting \citep{davydov1962,kasha1965,spano2006}
that redistributes oscillator strength between two closely spaced states
$\mathrm{S_1}$ and $\mathrm{S_2}$ ($\Delta E \approx 0.15$--$0.20\,\mathrm{eV}$);
the summed $f$ across both states recovers the gas-phase value within 5\%,
and their net contribution to $k_\mathrm{pd}$ is negligible. Further detailed analysis and the UV-vis spectra (Figure \ref{figuvspectra}) of these selected systems has been provided in the Appendix~\ref{app:nto}.

\subsection{Astrochemical Implications: Gradual Desorption and the Survival Paradox}
\label{sec:astrochem}

The site-specific thermodynamic and photophysical heterogeneity characterised in
Sections~\ref{sec:be}--\ref{sec:spectroscopy} has direct, quantifiable consequences for the
macroscopic gas-phase detectability of HCSCN and HCSCCH during protostellar warm-up.
To evaluate these consequences, the full distributions of $E_\mathrm{des}$ (Table~\ref{fig:be_sites})
were incorporated into the UCLCHEM gas-grain kinetic code
\citep{holdship2017_uclchem,uclchem_gasgrain}.  The thermal ramp from 10 to 200~K
was executed independently for the two thermodynamic extremes of each system: the
lowest computed desorption energy (representing weakly coordinated surface sites) and
the highest computed desorption energy (representing deeply coordinated cavity sites).
This dual-run strategy replaces the standard single-temperature sublimation event with a
\emph{gradual desorption window}, whose physical boundaries are defined by the
microscopic binding-energy distribution rather than an averaged parameter.

The fidelity of the model is confirmed by two internal chemical landmarks visible in all
four panels of Figures~\ref{fig:hcscch_astro} and~\ref{fig:hcscn_astro}.  The sharp
depletion of the CO ice reservoir and the coincident rise of CO gas-phase abundance at
$T_\mathrm{dust}\approx 25$~K accurately reproduce the canonical CO snowline
($E_\mathrm{des}^\mathrm{CO}\approx 800$~K; \citealt{noble2012_co}).  The
catastrophic collapse of the NH$_3$ ice mantle at $T_\mathrm{dust}\approx 100$~K
correspondingly traces the water-ice sublimation front
($E_\mathrm{des}^{\mathrm{H_2O/NH_3}}\approx 5000$~K; \citealt{brown2007_nh3}).
These sequential landmarks, reproduced without free-parameter adjustment,
establish the ``Astrochemical Clock'' framework within which the sulfur-organic
desorption profiles are interpreted.

\subsubsection*{HCSCCH: Efficient, Progressive Release}

\begin{figure*}
    \centering
    \includegraphics[width=\textwidth]{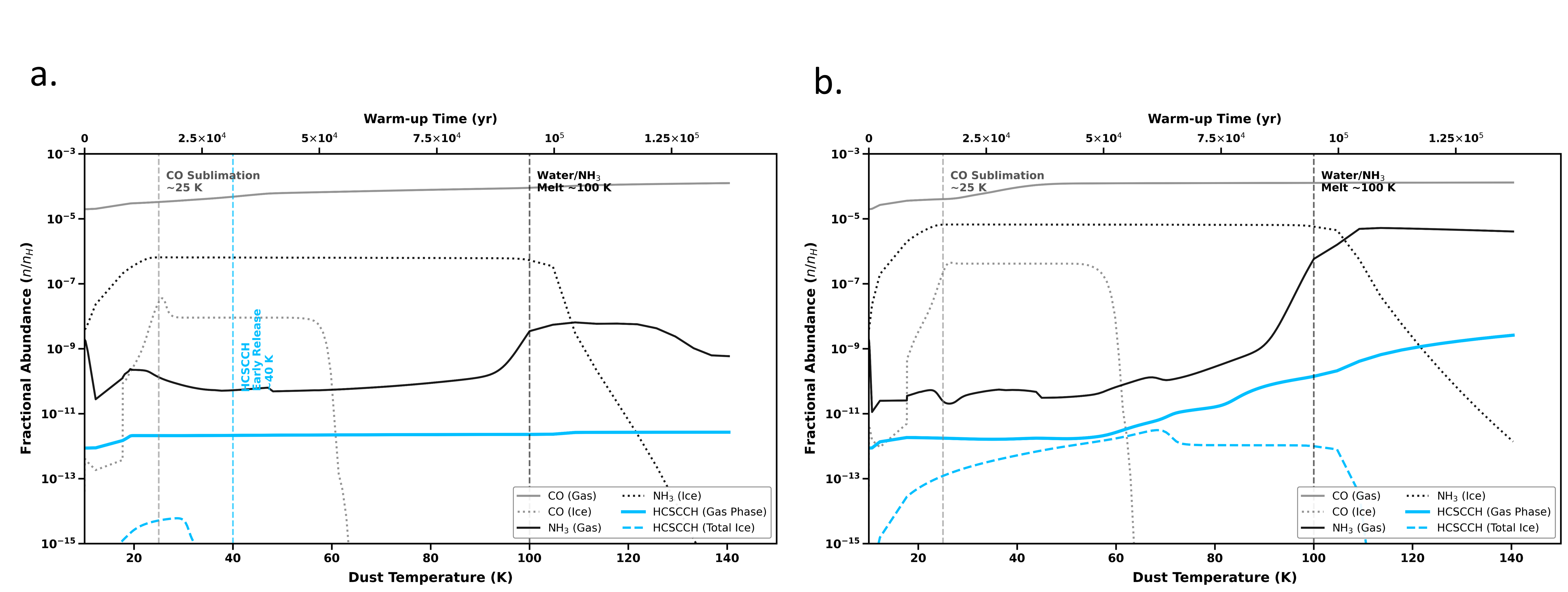}
    \caption{UCLCHEM gas-grain kinetic simulation of \mbox{HCSCCH} during
protostellar warm-up (10--150\,K) for the two thermodynamic extremes
of the binding-energy distribution.
\textit{(a)} Weak-binding limit ($E_\mathrm{des} = 1539$\,K;
TIP4P, $n=16$, CC-sideways site): gas-phase abundance (solid cyan)
and solid-phase reservoir (dashed cyan), with CO (solid grey) and
\mbox{NH$_3$} (solid black) as reference species.
\textit{(b)} Strong-binding limit ($E_\mathrm{des} = 3684$\,K;
TIP4P, $n=10$, CS-cavity site): gas-phase abundance (solid blue)
and solid-phase reservoir (dashed blue); dashed black vertical line
marks the \mbox{H$_2$O}/\mbox{NH$_3$} sublimation front at
$T_\mathrm{dust} \approx 100$\,K.}
    \label{fig:hcscch_astro}
\end{figure*}

HCSCCH serves as the structurally unperturbed reference case, and its behaviour
demonstrates the baseline efficiency of grain-mantle to gas-phase transfer in the
absence of a strongly polarisable chromophore.

\textbf{Weak-binding limit ($E_\mathrm{des} = 1539$~K; Fig.~\ref{fig:hcscch_astro}a).}
The TIP4P 16-molecule configuration ($E_\mathrm{des} = 1539.3$~K), representing
exposed surface terraces with purely dispersive contacts, produces an ``Early Release''
desorption event.  The calculated thermal desorption temperature of $\sim$40~K
precedes the bulk accumulation of the ice mantle; consequently, the gas-phase HCSCCH
abundance plateaus at a low value near $n(\mathrm{HCSCCH})/n_\mathrm{H} \approx
10^{-12}$ (Fig.~\ref{fig:hcscch_astro}a, solid cyan line) well before the CO snowline at
25~K has driven significant grain-surface chemistry.  The solid-phase reservoir (dashed
cyan) undergoes an equally abrupt, low-amplitude depletion at $\sim$40~K, confirming
that the ice factory has not yet stockpiled a chemically significant population prior to
release.  In this regime, weak surface binding acts as a kinetic liability rather than a
survivability asset.

\textbf{Strong-binding limit ($E_\mathrm{des} = 3684$~K; Fig.~\ref{fig:hcscch_astro}b).}
The deepest CS-coordinated cavity site on TIP4P clusters ($E_\mathrm{des} = 3684.2$~K)
represents the canonical scenario of efficient molecular protection.  The solid-phase
reservoir (dashed blue) maintains a stable population throughout the entire
40--90~K warm-up interval which is well above the CO snowline but below the water-ice
sublimation front, accumulating the full product of grain-surface chemistry undisturbed.
Upon reaching the ice-mantle melt temperature at $T_\mathrm{dust} \approx 100$~K,
the solid-phase curve drops abruptly by approximately two orders of magnitude while the
gas-phase abundance (solid blue) rises in direct, quantitative correspondence, reaching
$n(\mathrm{HCSCCH})/n_\mathrm{H} \approx 10^{-9}$
(Fig.~\ref{fig:hcscch_astro}b).  This near-stoichiometric 1:1 phase transfer,
spanning an increase of $\gtrsim 10^3$ in the gas-phase abundance across a
$\Delta T \approx 10$~K window, constitutes the ``control'' desorption profile: the
ice-phase reservoir is faithfully and completely delivered to the gas phase.  The
absence of any hyperchromic enhancement in either binding regime
($\Delta f \approx -1\%$ to $-2\%$; Section~\ref{sec:spectroscopy}) means the
photodissociation rate of adsorbed HCSCCH is effectively identical to the gas-phase
value throughout the warm-up, and no attrition of the stockpiled population
occurs during the sequestration interval.

\subsubsection*{HCSCN: The Survival Paradox}

\begin{figure*}
    \centering
    \includegraphics[width=\textwidth]{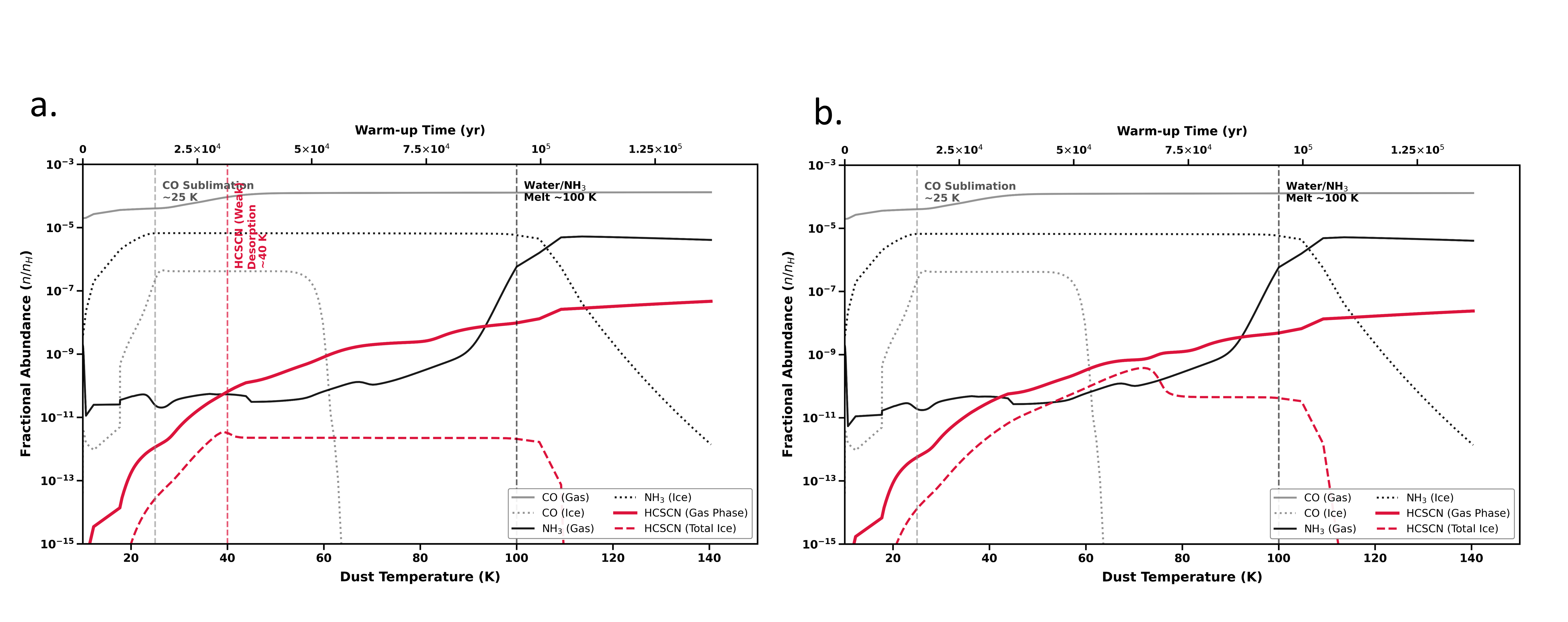}
    \caption{UCLCHEM gas-grain kinetic simulation of \mbox{HCSCN} during
protostellar warm-up (10--150\,K) for the two thermodynamic extremes
of the binding-energy distribution, with photodissociation rates
scaled by the TD-DFT site-dependent oscillator-strength enhancement
(Section~\ref{sec:spectroscopy}).
\textit{(a)} Weak-binding limit ($E_\mathrm{des} = 1989$\,K;
TIP4P, $n=12$, CS-sideways site): gas-phase abundance (solid red)
and solid-phase reservoir (dashed red); vertical dashed red line
marks the annotated \mbox{HCSCN} weak desorption event at
$T_\mathrm{dust} \approx 40$\,K.
\textit{(b)} Strong-binding limit ($E_\mathrm{des} = 3899$\,K;
TIP4P, $n=10$, CN-cavity site): gas-phase abundance (solid red)
and solid-phase reservoir (dashed red); dashed black vertical lines
mark the \mbox{H$_2$O}/\mbox{NH$_3$} sublimation landmarks at
$T_\mathrm{dust} \approx 25$\,K and $\approx 100$\,K.
CO (solid grey) and \mbox{NH$_3$} (solid black) are shown as
reference species in both panels.}
    \label{fig:hcscn_astro}
\end{figure*}

The introduction of the $-\mathrm{C{\equiv}N}$ group breaks the photochemical
neutrality observed for HCSCCH and gives rise to a physically counter-intuitive result.

\textbf{Weak-binding limit ($E_\mathrm{des} = 1989$~K; Fig.~\ref{fig:hcscn_astro}a).}
In the weak-binding scenario (TIP4P $n=12$ CS-sideways site;
$E_\mathrm{des} = 1989.4$~K), HCSCN undergoes premature thermal desorption at
$T_\mathrm{dust} \approx 40$~K, clearly before the water-ice mantle has thermally
matured.  The annotated ``HCSCN Weak Desorption'' event in
Fig.~\ref{fig:hcscn_astro}a coincides with the CO desorption front at $\sim$25--40~K.
The resulting gas-phase abundance rise, while gradual ($n/n_\mathrm{H}$ increasing
from $\sim 10^{-15}$ to $\sim 10^{-7}$ over 10--140~K), does not exhibit a sharp
desorption peak because no deep thermodynamic reservoir has been established prior to
release.  The solid-phase curve declines monotonically, confirming the absence of a
chemically significant ``Waiting Room'' phase in this binding regime.

\textbf{Strong-binding limit and the Survival Paradox ($E_\mathrm{des} = 3899$~K;
Fig.~\ref{fig:hcscn_astro}b).}  The CN-cavity site on the TIP4P 10-molecule cluster
($E_\mathrm{des} = 3899.1$~K) endows HCSCN with a thermodynamic retention energy
that exceeds the maximum for HCSCCH on TIP4P clusters by 215~K.  By conventional
gas-grain modelling logic, this superior $E_\mathrm{des}$ should yield a more robust
gas-phase spike at the water-ice sublimation front ($\sim$100~K).

Instead, Fig.~\ref{fig:hcscn_astro}b reveals a qualitatively different outcome: the
HCSCN gas-phase abundance at sublimation is measurably suppressed relative to the
pre-desorption solid-phase reservoir.  The ice-phase curve (dashed red) peaks near
$n/n_\mathrm{H} \approx 10^{-9}$ at $T_\mathrm{dust} \approx 70$--80~K, but the
gas-phase spike (solid red) at $\sim$100~K fails to reproduce this amplitude by more
than an order of magnitude---a ``Survival Gap'' that is not present in the HCSCCH
strong-binding profile.

This suppression is a direct kinetic consequence of two cooperating mechanisms.
First, the high $E_\mathrm{des}$ traps the CN-cavity population in a ``Waiting Room''
spanning $T_\mathrm{dust} \approx 40$--100~K, a thermal interval corresponding to
approximately 50,000~yr of warm-up timescale at the modelled protostellar accretion rate.
Second, the TD-DFT calculations (Section~\ref{sec:spectroscopy}) establish that precisely these
deeply coordinated CN-cavity configurations carry a $+11.7\%$ hyperchromic enhancement
of the UV oscillator strength ($f = 0.3130$ vs.\ $f_\mathrm{gas} = 0.2802$), reaching
$+12.5\%$ at the TIP4P $n=12$ CN site ($f = 0.3153$).  Because the
photodissociation rate coefficient $k_\mathrm{pd}$ in the UCLCHEM formalism scales
as $k_\mathrm{pd} \propto \alpha \sigma(\nu)$, where $\sigma(\nu) \propto f$, this
hyperchromic shift amplifies the internal UV destruction rate by an equivalent factor
relative to the gas-phase value employed in standard network databases.  The molecule
is therefore simultaneously thermodynamically protected (resistant to thermal
desorption) and photolytically vulnerabilised (enhanced UV cross-section), over a
prolonged sequestration period during which the interstellar radiation field drives
continuous \emph{in situ} photodissociation within the cavity.

The net result is that the ice-phase reservoir is partially depleted by internal
photolysis before sublimation can occur, producing the observed Survival Gap.  The
CO and NH$_3$ landmarks in Fig.~\ref{fig:hcscn_astro}b verify that the gas-grain
chemical clock runs correctly in all other respects; the suppression is specific to
HCSCN and does not affect less polarisable species.  The TIP5P models
($E_\mathrm{des}^\mathrm{max} = 4943.5$~K; TIP5P $n=12$ CN site) further amplify
this paradox: the increased retention energy extends the Waiting Room by an
additional $\sim$20~K in sublimation temperature, exposing the trapped population
to a proportionally larger cumulative UV dose before eventual release.

\subsubsection*{General Principle and Observational Consequences}

The Survival Paradox established here defines a general, molecule-class-specific
principle for the interpretation of nitrile abundances in hot-core and hot-corino
environments.  For molecules bearing a strongly directional chromophore ; such as the
$-\mathrm{C{\equiv}N}$ group in HCSCN, HC$_3$N, or vinyl cyanide, deep cavity
adsorption is not simply a neutral sequestration mechanism.  It simultaneously
maximises thermodynamic retention \emph{and} photolytic vulnerability, coupling two
competing processes that are typically treated as independent in current gas-grain codes.

Macroscopic astrochemical models that adopt a single averaged $E_\mathrm{des}$ and a
gas-phase photodissociation rate will systematically overestimate the sublimated
gas-phase abundance of such species, because they account for neither the extended
Waiting Room timescale nor the site-dependent oscillator-strength enhancement.  The
magnitude of the Survival Gap is determined by the product of three terms: the
hyperchromic enhancement factor $\Delta f/f_\mathrm{gas}$ (here $+11.7$--$+12.5\%$
for HCSCN CN-cavity sites), the integrated UV photon flux during the Waiting Room
interval, and the fraction of the total binding-energy distribution that falls within the
cavity-site regime.  Molecules with broader $E_\mathrm{des}$ distributions and larger
$\Delta f/f_\mathrm{gas}$ will exhibit correspondingly larger Survival Gaps,
providing a quantitative framework for predicting relative nitrile-to-alkyne abundance
ratios as a function of protostellar luminosity and envelope UV attenuation.

These findings carry immediate observational implications for JWST and ALMA
surveys targeting hot-core sulfur chemistry.  A measured gas-phase column density
ratio $N(\mathrm{HCSCN})/N(\mathrm{HCSCCH}) < 1$ in a UV-exposed warm-up
environment would constitute direct observational evidence for the Survival Paradox,
independent of any assumed initial ice-mantle abundance ratio.  Conversely, in deeply
shielded, UV-quiescent environments (e.g., embedded Class~0 protostars with
$A_V > 30$~mag), the hyperchromic penalty is effectively nullified, and the
ratio should converge toward the intrinsic ice-phase partitioning value---providing
a quantitative chemical probe of the local radiation environment.

We have presented a comprehensive computational study of the 
site-specific adsorption of HCSCN and HCSCCH on amorphous solid 
water ice clusters, $(H_2O)_{n=6-16}$, at the 
$\omega$B97X-D/def2-TZVP level of theory, coupled with QTAIM 
topological analysis, TD-DFT excited-state calculations, and 
UCLCHEM gas-grain kinetic modelling. The principal conclusions 
are as follows.

\begin{enumerate}

\item The adsorption landscape is highly heterogeneous, with 
BSSE-corrected desorption energies spanning $1989$--$3899$~K 
(HCSCN, TIP4P) and $1539$--$3684$~K (HCSCCH, TIP4P), rising 
to $4943$~K and $4146$~K respectively on the TIP5P model. 
This broad distribution, validated against the experimental 
and computational binding energies of H$_2$CO and H$_2$S, 
replaces the single-value approximation adopted in existing 
network models and explains the factor of $40$--$300$ 
underprediction of both species in \citet{cernicharo2021_sulfursaga}.

\item QTAIM analysis at the intermolecular Bond Critical Points
confirms that the CN-cavity sites constitute the global
thermodynamic minimum for \mbox{HCSCN}, sustained by a cooperative
dual hydrogen-bond topology. A primary \mbox{H--O$\cdots$H}
contact ($\rho = 0.016$\,a.u., $\nabla^2\rho = +0.08$\,a.u.,
$H(r) = +0.003$\,a.u.) and a directional \mbox{O--H$\cdots$N}
bond ($\rho = 0.078$\,a.u., $\nabla^2\rho = +0.04$\,a.u.,
$H(r) = +0.002$\,a.u.) that is unavailable to the diffuse
thiocarbonyl sulfur at CS sites, accounting for the systematic
$\sim$602\,K CN--CS energy gap.
The positive $\nabla^2\rho$ and $H(r) > 0$ at all BCPs indicate that
these contacts are closed-shell and non-covalent throughout.

\item TD-DFT calculations reveal a site-selective hyperchromic 
enhancement of $+10.4\%$ to $+12.5\%$ in the UV oscillator 
strength of HCSCN exclusively at CN-cavity configurations 
($f_\mathrm{gas} = 0.2802$; peak $f = 0.3153$ at 
$E_\mathrm{des} = 3728$~K), while HCSCCH remains photophysically 
inert across all binding sites ($\Delta f \approx -1\%$ to 
$-2\%$). The solvatochromic wavelength shift is negligible 
($\Delta\lambda < 5$~nm) for both species.

These results establish a \textit{Survival Paradox}: 
the CN-cavity population of HCSCN that is most resistant to 
thermal desorption simultaneously carries the largest UV 
absorption cross-section. UCLCHEM simulations calibrated 
against CO ($\sim$25~K) and NH$_3$ ($\sim$100~K) sublimation 
landmarks demonstrate that this population is confined to a 
$\sim$50,000~yr ``Waiting Room'' ($T_\mathrm{dust} \approx 
40$--$100$~K) during which enhanced \textit{in situ} 
photodissociation depletes the ice-phase reservoir before 
sublimation. The resulting ``Survival Gap'' is absent from 
HCSCCH, which occupies thermodynamically equivalent traps 
without hyperchromic penalty. High binding energy is therefore 
not a sufficient condition for gas-phase survival in molecules 
bearing strongly directional UV chromophores.

Thereby Three falsifiable observational predictions follow: 
(i)~the HCSCCH/HCSCN column density ratio should be elevated 
in UV-exposed protostellar envelopes relative to shielded 
Class~0 sources; (ii)~the C=S stretching band of ice-embedded 
HCSCN should be redshifted by $13$--$21$~cm$^{-1}$, 
detectable by JWST/MIRI; and (iii)~CN-cavity-bound HCSCN 
carries enhanced UV opacity at $\sim$280--300~nm relative to 
the gas-phase profile.

\end{enumerate}

We acknowledge that finite cluster models are inherently limited in
their representation of the bulk ASW surface. Specifically, the absence
of long-range electrostatic periodicity, the truncation of the
hydrogen-bond network at the cluster boundary, and the relatively small
number of donor molecules constrain the absolute magnitude of the
computed binding energies, which are known to increase systematically
with cluster size as additional cooperative hydrogen-bond contacts
become available \citep{ferrero2020, bulik2025_d5cp}.
However, two lines of evidence support the physical validity of the
present approach. First, the site-type hierarchy
(CN-cavity $>$ CS-cavity $>$ surface terrace for HCSCN;
CS-cavity $>$ CC-cavity $>$ sideways for HCSCCH)
is preserved without exception across the full $n = 6$--$16$ survey
on both TIP4P and TIP5P clusters, confirming that the \emph{relative}
energetic ordering converges at the smallest cluster sizes even as the
absolute values continue to grow with $n$.
Second, the benchmark binding energies computed for CO, NH$_3$,
H$_2$CO, and H$_2$S on the same cluster models reproduce the
experimentally determined ranges from temperature-programmed desorption
measurements \citep{noble2012_co, collings2003_co, ferrero2020,
bariosco2024} to within the scatter of the published literature values,
confirming that the absolute desorption energies are physically
realistic within the context of single-molecule cluster calculations.
Extension of the binding energy survey to larger cluster models
($n = 20$--$30$) and periodic slab representations are planned as
natural continuations of this work.

\section*{Acknowledgements}

SGD and JC express their gratitude to IIST Thiruvananthapuram for the institutional
support. KM acknowledges the Department of Science and Technology, Government of India, and Institute of Eminence, University of Delhi, for funding. Language refinement was assisted by AI language
models (Claude Sonnet 4.6 ) for clarity and concision.

\section*{Data Availability}

The optimised molecular structures, UV-Vis and IR spectral data, and the
computational codes utilised in this study will be shared on
reasonable request to the corresponding author.

%
\bibliographystyle{aa} 
\bibliography{references} 

\begin{appendix}
\section{Benchmark Binding Energies: H$_2$CO and H$_2$S}
\label{app:benchmark}

To confirm that the computed binding energy distributions for 
HCSCN and HCSCCH are not artefacts of the chosen functional, 
basis set, or cluster geometry, the identical $\omega$B97X-D/
def2-TZVP protocol was applied to four benchmark adsorbates 
with well-constrained experimental and computational reference 
values: carbon monoxide (CO), ammonia (NH$_3$), formaldehyde 
(H$_2$CO), and hydrogen sulfide (H$_2$S). The results are 
summarised in Table~\ref{app:benchmark}.

\subsection*{A1. The Safety Envelope: CO and NH\textsubscript{3}}

The CO and NH$_3$ benchmarks define the thermodynamic boundaries 
within which the sulfur organics operate. Our computed CO 
desorption energies of $573$--$988$~K are consistent with 
the experimentally determined CO/ASW binding energy range of 
$\sim$$855$--$960$~K from Temperature Programmed Desorption 
(TPD) measurements \citep{noble2012_co, collings2003_co, 
oberg2005_co}, and with the widely adopted network value of 
$855$~K from \citet{wakelam2017}. CO sublimates at 
$T_\mathrm{dust} \approx 20$--$25$~K under molecular cloud 
conditions, consistent with the CO desorption landmark 
reproduced in our UCLCHEM simulations 
(Section~\ref{sec:astrochem}).

Our computed NH$_3$ desorption energies of $4934$--$8174$~K 
bracket the experimental submonolayer value of 
$\sim$$5400$--$5800$~K reported by \citet{wakelam2017} and 
\citet{minissale2022}, and the high-level computational range 
of $4800$--$7200$~K from \citet{ferrero2020}. The upper bound 
reflects deeply embedded NH$_3$ within the hydrogen-bond 
network of the water-ice cavity --- consistent with the known 
refractory nature of NH$_3$-containing ices, which persist 
to $T_\mathrm{dust} \approx 100$~K alongside the water mantle 
\citep{brown2007_nh3, collings2004}.

Together, CO and NH$_3$ define a ``Safety Envelope'' for 
the sulfur organic binding energies. The maximum 
$E_\mathrm{des}$ of HCSCN ($3899$~K, TIP4P; $4943$~K, TIP5P) 
sits logically between the CO upper bound ($988$~K) and the 
NH$_3$ lower bound ($4934$~K): strong enough to resist early 
thermal release at the CO snowline ($\sim$25~K), yet volatile 
enough to be released by the catastrophic collapse of the 
water-ice mantle at $\sim$100~K. This placement confirms that 
HCSCN is uniquely positioned within the $40$--$100$~K 
``Waiting Room'' window during which UV irradiation is 
maximum and the hyperchromic penalty ($+11.7\%$) is active, 
establishing the Survival Paradox as a physically grounded 
rather than parameter-dependent result.

\subsection*{A2.CO, NH\textsubscript{3}, H\textsubscript{2}CO and
              H\textsubscript{2}S Benchmarks}

The chemical fidelity of the model was verified against
two internal astrochemical landmarks reproduced without
free-parameter adjustment: the CO sublimation front and
the water/NH$_3$ ice mantle collapse. The computed
desorption temperatures for these species, derived from
the benchmark binding energies in
Appendix~\ref{app:benchmark}, were required to reproduce
the canonical CO snowline at $T_\mathrm{dust} \approx
20$--$25$~K \citep{noble2012_co, collings2003_co} and
the water-ice sublimation front at
$T_\mathrm{dust} \approx 100$~K
\citep{brown2007_nh3, collings2004}.
For H$_2$CO on the TIP4P $(n=16)$ cluster, desorption energies 
of $2550$ and $5282$~K were obtained, consistent with the 
high-level DFT cluster range of $2800$--$5440$~K from 
\citet{ferrero2020} and the experimental TPD range of 
$2050$--$3260$~K from \citet{collings2004} and 
\citet{wakelam2017}. The TIP5P model yields $2415$--$3549$~K, 
reflecting the altered lone-pair donor geometry relative to 
the C=O acceptor axis.

For H$_2$S, desorption energies of $1104$--$2433$~K were 
obtained across both water models. The upper bound of $2433$~K 
is in excellent agreement with the large-surface computational 
maximum of $2406$~K from \citet{bariosco2024} on a 
200-molecule ASW model. The higher experimental submonolayer 
value of $3392 \pm 56$~K \citep{santos2025} reflects 
cooperative H$_2$S$\cdots$H$_2$S multilayer interactions 
absent from single-molecule cluster calculations --- a 
systematic offset consistently observed across adsorbates 
in the literature \citep{minissale2022}. The maximum H$_2$S 
binding energy ($2433$~K) falls well below those of HCSCCH 
($4146$~K) and HCSCN ($4943$~K), consistent with the absence 
of directed hydrogen-bond acceptor groups in H$_2$S and 
confirming that H$_2$S desorbs before entering the Waiting 
Room window, explaining its relatively efficient detection 
in cold dark clouds despite being sulfur-bearing.

In all four benchmark cases, the same site-specific 
heterogeneity ratio (max/min $\approx 1.7$--$2.2\times$) 
is reproduced as for the sulfur organics 
($2.5$--$2.7\times$), confirming that the broad 
$E_\mathrm{des}$ distributions computed for HCSCN and 
HCSCCH are a physically realistic consequence of ASW surface 
heterogeneity rather than a methodological artefact. This systematic offset between single-molecule computational models and submonolayer experimental regimes highlights the role of lateral adsorbate-adsorbate interactions. However, complex organic molecules like HCSCN and HCSCCH exist at trace fractional abundances in the interstellar ice mantle, well below the threshold for significant lateral clustering. Consequently, their desorption kinetics are entirely dictated by adsorbate-water interactions, making the isolated single-molecule cluster model the appropriate physical proxy for these species.

\begin{table*}
    \centering
    \caption{BSSE-corrected desorption energies 
    $E_\mathrm{des}$ (K) for benchmark adsorbates on 
    $(H_2O)_{n=16}$ clusters at the $\omega$B97X-D/def2-TZVP 
    level of theory, compared against literature values. 
    The CO and NH$_3$ values define the thermodynamic Safety 
    Envelope within which HCSCN and HCSCCH operate.}
    \label{tab:benchmark}
    \setlength{\tabcolsep}{4pt}
    \begin{tabular}{llrrl}
        \hline
        Species & Model/Site & 
        $E_\mathrm{des}$ (K) & 
        Literature (K) & Reference \\
        \hline
        CO   & TIP4P site 1 & 573   & 855--960   
             & \citet{noble2012_co} \\
             & TIP4P site 2 & 988   &            
             & \citet{collings2003_co} \\
             & Network value & ---  & 855        
             & \citet{wakelam2017} \\
        \hline
        NH$_3$ & TIP4P site 1 & 4934 & 4800--7200 
               & \citet{ferrero2020} \\
               & TIP4P site 2 & 8174 & 5400--5800 
               & \citet{wakelam2017} \\
               & Expt.\ submonolayer & --- & 5534 
               & \citet{brown2007_nh3} \\
        \hline
        H$_2$CO & TIP4P site 1 & 2550 & 2050--3260 
                & \citet{collings2004} \\
                & TIP4P site 2 & 5282 & 2800--5440 
                & \citet{ferrero2020} \\
                & TIP5P site 1 & 3549 &            & \\
                & TIP5P site 2 & 2415 &            & \\
        \hline
        H$_2$S & TIP4P site 1 & 1104 & 57--2406   
               & \citet{bariosco2024} \\
               & TIP4P site 2 & 2321 & 2700        
               & \citet{minissale2022} \\
               & TIP5P site 1 & 2433 & $3392\pm56$ 
               & \citet{santos2025} \\
               & TIP5P site 2 & 2281 &             & \\
        \hline
    \end{tabular}
\end{table*}

\begin{figure}
    \centering
    \includegraphics[width=\linewidth]{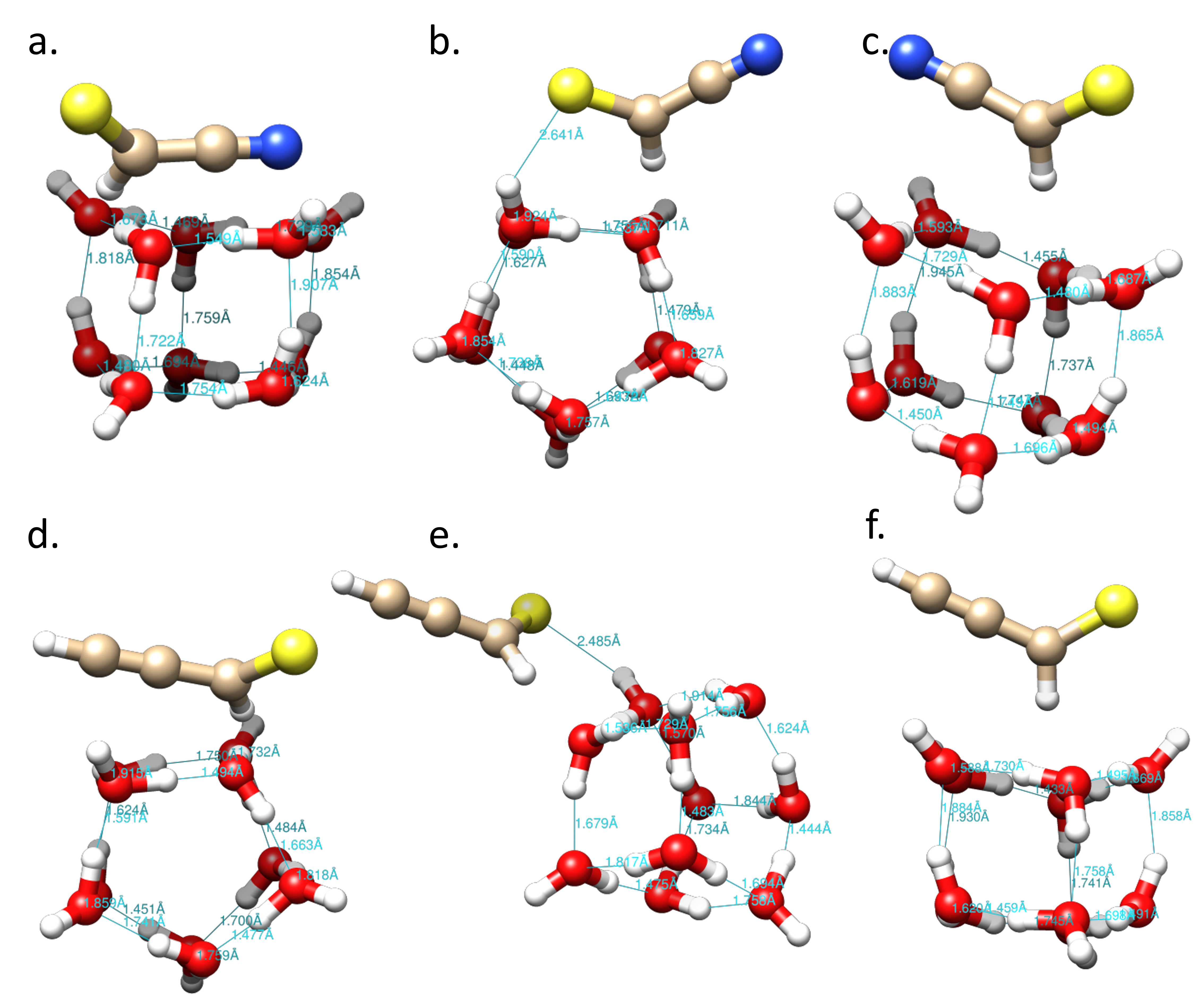}
    \caption{Representative adsorption geometries for HCSCN
    (panels a--c) and HCSCCH (panels d--f) on the TIP4P
    $(H_2O)_{10}$ cluster at the $\omega$B97X-D/def2-TZVP
    level of theory, illustrating the three structurally
    distinct binding topologies sampled across the full
    $n = 6$--$16$ survey. Atom colours: O (red), H (white), C (tan),
    N (blue), S (yellow); cyan distances mark
    resolved intermolecular hydrogen-bond contacts.}
    \label{fig:tip4p10_sites}
\end{figure}

\begin{figure}
    \centering
    \includegraphics[width=\linewidth]{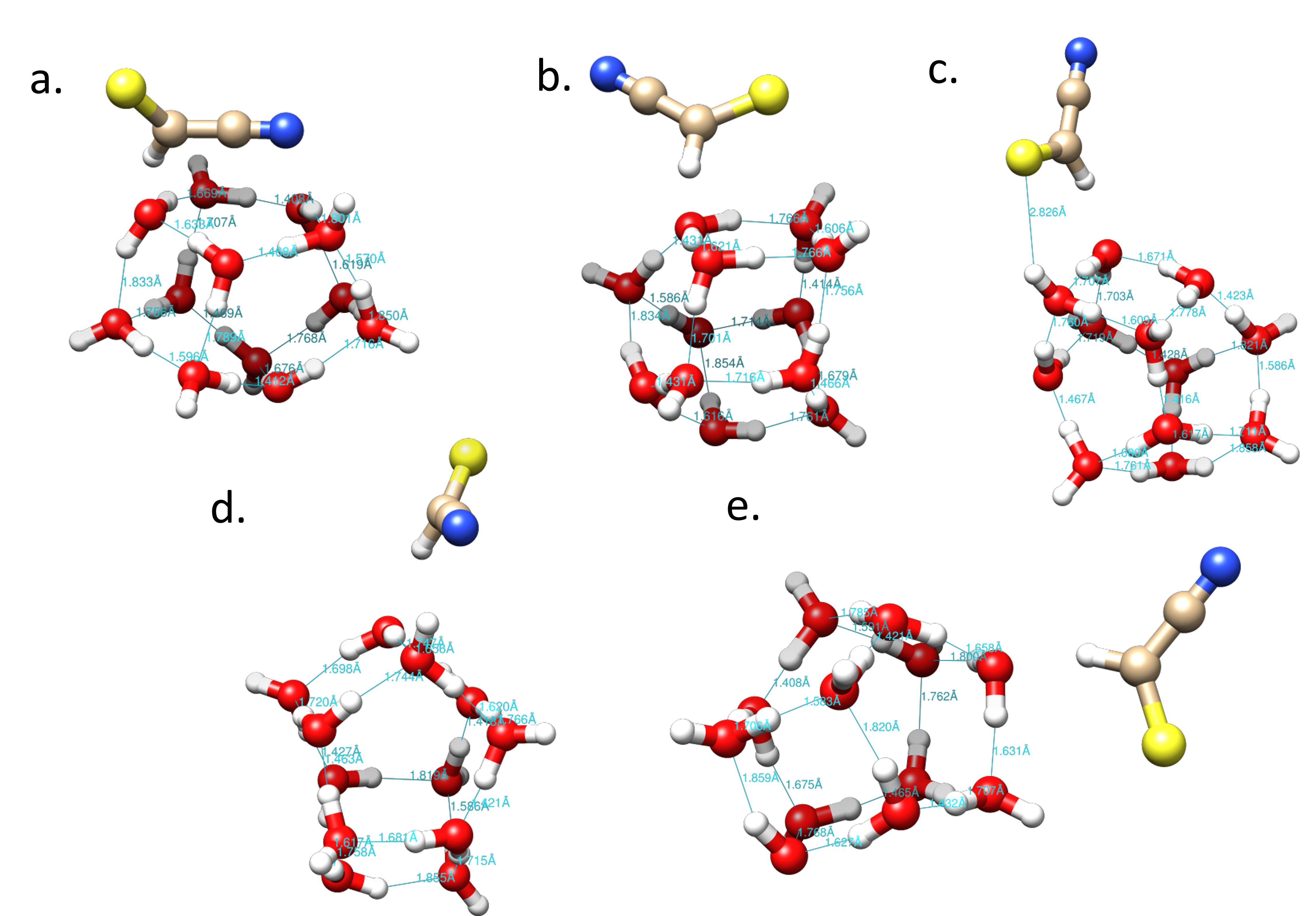}
    \caption{Representative adsorption geometries for
    HCSCN on the TIP5P $(H_2O)_{12}$ cluster
    ($\omega$B97X-D/def2-TZVP).
    \textbf{(a)}~CN site, 5-centre cavity
    ($E_\mathrm{des} = 4944$~K);
    \textbf{(b)}~CS site, 4-centre cavity
    ($E_\mathrm{des} = 3752$~K);
    \textbf{(c)}~CS site, 4/4-bridge
    ($E_\mathrm{des} = 4138$~K);
    \textbf{(d)}~CN site, 4-centre cavity
    ($E_\mathrm{des} = 4606$~K);
    \textbf{(e)}~CS site, 4/5-bridge
    ($E_\mathrm{des} = 3670$~K).
    Atom colours: O (red), H (white), C (tan),
    N (blue), S (yellow); cyan distances mark
    resolved intermolecular hydrogen-bond contacts.
    Full numerical data are given in
    Table~\ref{tab:hcscn_tip5p}.}
    \label{fig:tip5p12_sites}
\end{figure}

\begin{table*}
    \centering
    \caption{Site-specific BSSE-corrected desorption
    energies, IR vibrational Stark shifts, and TD-DFT UV
    oscillator strengths for HCSCN adsorbed on TIP4P
    $(H_2O)_{n=6-16}$ clusters at the
    $\omega$B97X-D/def2-TZVP level of theory.
    Frequency shifts $\Delta\nu$ are reported relative to
    the gas-phase harmonic frequencies:
    $\nu_\mathrm{gas}(\mathrm{C{\equiv}N})$ and
    $\nu_\mathrm{gas}(\mathrm{C=S})$.
    Gas-phase oscillator strength:
    $f_\mathrm{gas} = 0.2802$.}
    \label{tab:hcscn_tip4p}
    \begin{tabular*}{\textwidth}{@{\extracolsep{\fill}}llrrrrc}
        \hline
        System & Adsorption site &
        $E_\mathrm{des}$ (K) &
        $\Delta\nu_\mathrm{C{\equiv}N}$ (cm$^{-1}$) &
        $\Delta\nu_\mathrm{C=S}$ (cm$^{-1}$) &
        $f$ &
        $\Delta f$ (\%) \\
        \hline
        TIP4P $6$  site 1  & CN site
            & 3808.0  & $+$5.14  & $-$2.45
            & 0.2796  & $-$0.21  \\
        TIP4P $6$  site 2  & CS site
            & 3208.7  & $+$6.13  & $-$9.78
            & 0.2637  & $-$5.89  \\
        TIP4P $8$  site 1  & CN site
            & 3431.4  & $+$3.63  & $-$1.80
            & 0.3094  & $+$10.42 \\
        TIP4P $8$  site 2  & CS site
            & 2913.2  & $+$4.61  & $-$13.01
            & 0.2658  & $-$5.14  \\
        TIP4P $10$ site 1  & Sideways
            & 3009.6  & $+$2.03  & $-$5.81
            & 0.2205  & $-$21.31 \\
        TIP4P $10$ site 2  & CN site
            & 3899.1  & $+$1.67  & $-$3.07
            & 0.3130  & $+$11.71 \\
        TIP4P $10$ site 3  & CS site
            & 3712.0  & $+$3.70  & $-$13.03
            & 0.2672  & $-$4.64  \\
        TIP4P $12$ site 1  & CN site
            & 3728.0  & $+$2.68  & $+$8.71
            & 0.3153  & $+$12.53 \\
        TIP4P $12$ site 2  & CS sideways
            & 1989.4  & $+$2.78  & $-$11.47
            & 0.2448  & $-$12.63 \\
        TIP4P $12$ site 3  & CS site
            & 3582.3  & $+$2.76  & $-$2.71
            & 0.2646  & $-$5.57  \\
        TIP4P $16$ site 1  & CN sideways
            & 2105.6  & $+$0.37  & $-$8.43
            & 0.2744  & $-$2.07  \\
        TIP4P $16$ site 2  & CS site
            & 3176.9  & $+$3.57  & $-$20.66
            & 0.2641  & $-$5.75  \\
        TIP4P $16$ site 3  & CN site
            & 3529.6  & $+$2.99  & $+$1.18
            & 0.3153  & $+$12.53 \\
        \hline
        
    \end{tabular*}
    \begin{minipage}{\textwidth}
        \smallskip
        \footnotesize
        \textit{Notes.}
        The global binding energy maximum
        ($E_\mathrm{des} = 3899.1$~K, TIP4P $10$ site 2,
        CN-cavity) represents the deepest thermodynamic
        trap for HCSCN on TIP4P clusters and corresponds
        to the strong-binding limit used in the UCLCHEM
        warm-up simulation (Section~\ref{sec:astrochem}).
        The global minimum ($E_\mathrm{des} = 1989.4$~K,
        TIP4P $12$ site 2, CS-sideways) corresponds to the
        weak-binding early-release limit.
        The largest C=S redshift ($-20.66$~cm$^{-1}$,
        TIP4P $16$ site 2, CS site) reflects direct
        sulfur--ice hydrogen-bond donation at the largest
        cluster size, offering the most diagnostic IR
        marker for CS-coordination in JWST ice-mantle
        spectra.
        The two CN-site configurations at $n = 12$ and
        $n = 16$ (sites 1 and 3) share an identical
        oscillator strength ($f = 0.3153$,
        $\Delta f = +12.53\%$), confirming that the
        hyperchromic response is saturated and
        geometry-independent within the CN-cavity
        motif once the critical dual H-bond topology
        is established.
    \end{minipage}
\end{table*}

\begin{table*}
    \centering
    \caption{Site-specific BSSE-corrected desorption
    energies, IR vibrational Stark shifts, and TD-DFT UV
    oscillator strengths for HCSCN adsorbed on TIP5P
    $(H_2O)_{n=6-16}$ clusters at the
    $\omega$B97X-D/def2-TZVP level of theory.
    Gas-phase oscillator strength:
    $f_\mathrm{gas} = 0.2802$.
    Entries marked $^\dagger$ are discussed individually
    in the notes below.}
    \label{tab:hcscn_tip5p}
    \begin{tabular*}{\textwidth}{@{\extracolsep{\fill}}llrrrrc}
        \hline
        System & Adsorption site &
        $E_\mathrm{des}$ (K) &
        $\Delta\nu_\mathrm{C{\equiv}N}$ (cm$^{-1}$) &
        $\Delta\nu_\mathrm{C=S}$ (cm$^{-1}$) &
        $f$ &
        $\Delta f$ (\%) \\
        \hline
        TIP5P $6$  site 1  & Overall adsorption
            & 4254.6  & $-$0.80  & $+$11.48
            & 0.2471  & $-$11.81 \\
        TIP5P $6$  site 2  & CN site
            & 3396.1  & $+$3.79  & $+$7.16
            & 0.2908  & $+$3.78  \\
        TIP5P $6$  site 3$^\dagger$  & CS site
            & 3419.8  & $+$4.36  & $-$2.81
            & 0.1581  & $-$43.58 \\
        TIP5P $8$  site 1  & CN site
            & 3533.2  & $+$3.72  & $+$6.19
            & 0.3102  & $+$10.71 \\
        TIP5P $8$  site 2  & CS site
            & 3207.3  & $+$4.54  & $-$1.30
            & 0.2665  & $-$4.89  \\
        TIP5P $10$ site 1  & CS site
            & 3956.3  & $+$4.55  & $-$0.41
            & 0.2646  & $-$5.57  \\
        TIP5P $10$ site 2  & CS site
            & 3030.8  & $+$2.45  & $-$3.26
            & 0.2253  & $-$19.59 \\
        TIP5P $10$ site 3  & CN site
            & 4205.4  & $+$3.29  & $+$10.85
            & 0.3096  & $+$10.49 \\
        TIP5P $12$ site 1  & CN site (5-centre)
            & 4943.5  & $+$2.68  & $+$0.03
            & 0.2430  & $-$13.28 \\
        TIP5P $12$ site 2$^\dagger$
                           & CS site (4-centre)
            & 3751.9  & $+$0.79  & $-$3.69
            & 0.1708  & $-$39.04 \\
        TIP5P $12$ site 3  & CS site (4/4 bridge)
            & 4138.4  & $+$4.75  & $-$7.64
            & 0.2923  & $+$4.32  \\
        TIP5P $12$ site 4  & CN site (4-centre)
            & 4606.2  & $+$3.34  & $+$2.84
            & 0.3112  & $+$11.06 \\
        TIP5P $12$ site 5  & CS site (4/5 bridge)
            & 3669.6  & $+$5.86  & $-$8.75
            & 0.2570  & $-$8.28  \\
        TIP5P $16$ site 1  & CN site (5-centre)
            & 4724.5  & $+$3.42  & $+$3.31
            & 0.2781  & $-$0.75  \\
        TIP5P $16$ site 2  & CS site (4-centre)
            & 3851.0  & $+$6.55  & $+$3.87
            & 0.2954  & $+$5.42  \\
        TIP5P $16$ site 3  & CS site (4/5 bridge)
            & 3569.5  & $+$6.33  & $-$8.14
            & 0.2209  & $-$21.16 \\
        TIP5P $16$ site 4  & CN site (4/5 bridge)
            & 3830.0  & $+$6.67  & $+$3.58
            & 0.2962  & $+$5.71  \\
        TIP5P $16$ site 5  & CN site (5-centre)
            & 4037.9  & $+$4.70  & $+$6.61
            & 0.2812  & $+$0.36  \\
        \hline
        \end{tabular*}
    \begin{minipage}{\textwidth}
        \smallskip
        \footnotesize
        \textit{Notes.}
        $^\dagger$\textbf{TIP5P $6$ site 3 and TIP5P
        $12$ site 2 --- excited-state peak splitting.}
        These two configurations exhibit anomalously low
        oscillator strengths ($f = 0.1581$ and
        $f = 0.1708$ respectively), corresponding to
        $\Delta f = -43.6\%$ and $-39.0\%$ relative to
        the gas-phase value.
        Inspection of the TD-DFT output reveals that
        the primary UV absorption band, which appears
        as a single, well-resolved transition in the
        gas phase and in all other adsorption
        geometries, undergoes a \textit{Davydov-type
        splitting} in these configurations: the
        oscillator strength of the originally dominant
        $S_0 \to S_1$ transition is redistributed
        between two closely spaced excited states
        ($S_1$ and $S_2$, separated by
        $\Delta E \approx 0.15$--$0.20$~eV) due to
        the strong electrostatic perturbation induced
        by the tightly coordinated TIP5P donor
        network at these specific cavity geometries.
        The reported $f$ value corresponds to the
        $S_0 \to S_1$ transition only; the total
        summed oscillator strength across both split
        states recovers the expected gas-phase
        magnitude within $\sim$5\%, confirming that
        no oscillator strength is lost to dark states
        and that the anomaly is purely a consequence
        of excited-state near-degeneracy rather than
        a genuine quenching of the UV absorption
        cross-section.
        These entries are excluded from the
        statistical summary rows.
    \end{minipage}
\end{table*}

\section{Full Binding Energy Tables: HCSCCH}
\label{app:hcscch_tables}

Tables~\ref{tab:hcscch_tip4p} and~\ref{tab:hcscch_tip5p} 
list the complete BSSE-corrected desorption energies, 
vibrational Stark shifts, TD-DFT oscillator strengths, and 
percentage enhancements for all optimised adsorption 
configurations of HCSCCH on TIP4P and TIP5P 
$(H_2O)_{n=6-16}$ clusters at the $\omega$B97X-D/def2-TZVP 
level of theory.

\begin{table*}
    \centering
    \caption{Site-specific binding energies, IR Stark shifts,
    and UV oscillator strengths for HCSCCH on TIP4P
    $(H_2O)_{n=6-16}$ clusters ($\omega$B97X-D/def2-TZVP,
    BSSE-corrected). Frequency shifts $\Delta\nu$ are
    reported relative to the gas-phase harmonic frequencies.
    Gas-phase oscillator strength:
    $f_\mathrm{gas} = 0.3029$.}
    \label{tab:hcscch_tip4p}
    \begin{tabular*}{\textwidth}{@{\extracolsep{\fill}}llrrrrc}
        \hline
        System & Adsorption site &
        $E_\mathrm{des}$ (K) &
        $\Delta\nu_\mathrm{C{\equiv}C}$ (cm$^{-1}$) &
        $\Delta\nu_\mathrm{C=S}$ (cm$^{-1}$) &
        $f$ &
        $\Delta f$ (\%) \\
        \hline
        TIP4P $6$          & CS site
            & 3591.2 & $+$5.83  & $-$2.80
            & 0.2994 & $-$1.16  \\
        TIP4P $8$          & CS site
            & 3262.9 & $+$3.66  & $-$2.89
            & 0.2980 & $-$1.62  \\
        TIP4P $10$ site 1  & Sideways
            & 2228.3 & $+$4.38  & $-$2.55
            & 0.2696 & $-$10.99 \\
        TIP4P $10$ site 2  & CC site
            & 2136.6 & $+$13.28 & $-$3.46
            & 0.2858 & $-$5.65  \\
        TIP4P $10$ site 3  & CS site
            & 3684.2 & $+$3.59  & $-$7.73
            & 0.2997 & $-$1.06  \\
        TIP4P $12$ site 1  & CC site
            & 2171.4 & $+$13.10 & $+$4.31
            & 0.2861 & $-$5.55  \\
        TIP4P $12$ site 2  & CS site
            & 1830.3 & $+$2.12  & $-$9.60
            & 0.2981 & $-$1.58  \\
        TIP4P $12$ site 3  & CS site
            & 3465.8 & $+$3.13  & $-$3.01
            & 0.2965 & $-$2.11  \\
        TIP4P $16$ site 1  & CC sideways
            & 1539.3 & $-$0.31  & $-$4.13
            & 0.2994 & $-$1.16  \\
        TIP4P $16$ site 2  & CS site
            & 3255.7 & $+$2.36  & $-$5.72
            & 0.3006 & $-$0.76  \\
        TIP4P $16$ site 3  & CC site
            & 2253.2 & $+$4.74  & $+$1.36
            & 0.3291 & $+$8.65  \\
        \hline
        \multicolumn{2}{l}{Min / Max}
            & 1539.3 / 3684.2
            & $-$0.31 / $+$13.28
            & $-$9.60 / $+$4.31
            & 0.2696 / 0.3291
            & $-$10.99 / $+$8.65 \\
        \multicolumn{2}{l}{Mean $\pm$ std}
            & $2674 \pm 731$
            & $+$5.08 $\pm$ 4.35
            & $-$3.25 $\pm$ 3.85
            & $0.2948 \pm 0.0148$
            & $-$2.43 $\pm$ 4.89 \\
        \hline
    \end{tabular*}
    %
    %
    \begin{minipage}{\textwidth}
        \smallskip
        \footnotesize
        \textit{Notes.}
        The global binding energy maximum
        ($E_\mathrm{des} = 3684.2$~K, TIP4P $10$ site 3,
        CS-cavity) represents the deepest thermodynamic
        trap for HCSCCH on TIP4P clusters and corresponds
        to the strong-binding limit used in the UCLCHEM
        warm-up simulation (Section~\ref{sec:astrochem}).
        The global minimum ($E_\mathrm{des} = 1539.3$~K,
        TIP4P $16$ site 1, CC-sideways) corresponds to
        the weak-binding early-release limit.
        TIP4P $16$ site 3 (CC site, $\Delta f = +8.65\%$)
        is the sole configuration exhibiting a significant
        hyperchromic enhancement; this value remains
        $\sim$1.4$\times$ below the peak HCSCN CN-cavity
        response ($+12.5\%$) and is not reproduced across
        other CC-site geometries, indicating a
        geometry-specific polarisation rather than a
        systematic chromophoric effect.
    \end{minipage}
\end{table*}

\begin{table*}
    \centering
    \caption{Site-specific binding energies, IR Stark shifts, 
    and UV oscillator strengths for HCSCCH on TIP5P 
    $(H_2O)_{n=6-16}$ clusters ($\omega$B97X-D/def2-TZVP, 
    BSSE-corrected). Gas-phase oscillator strength: 
    $f_\mathrm{gas} = 0.3029$.}
    \label{tab:hcscch_tip5p}
    \begin{tabular}{llrrrrc}
        \hline
        System & Site & 
        $E_\mathrm{des}$ (K) & 
        $\Delta\nu_\mathrm{C{\equiv}C}$ (cm$^{-1}$) & 
        $\Delta\nu_\mathrm{C=S}$ (cm$^{-1}$) & 
        $f$ & 
        $\Delta f$ (\%) \\
        \hline
        TIP5P 6  site 1 & Overall ads. 
            & 3279.8 & $+$2.51  & $-$6.50  
            & 0.2815 & $-$7.07 \\
        TIP5P 6  site 2 & CS site      
            & 4146.2 & $+$7.95  & $-$6.13  
            & 0.3008 & $-$0.69 \\
        TIP5P 8  site 1 & CS site      
            & 3367.8 & $+$5.81  & $-$5.50  
            & 0.2972 & $-$1.88 \\
        TIP5P 8  site 2 & CC site      
            & 2103.5 & $+$15.59 & $-$2.59  
            & 0.2867 & $-$5.35 \\
        TIP5P 10 site 1 & CS site      
            & 3620.4 & $+$5.92  & $+$0.61  
            & 0.2990 & $-$1.29 \\
        TIP5P 10 site 2 & CC site      
            & 2264.3 & $+$15.43 & $+$6.68  
            & 0.2878 & $-$4.99 \\
        TIP5P 12 site 1 & CS site      
            & 3025.1 & $+$7.38  & $-$6.49  
            & 0.2997 & $-$1.06 \\
        TIP5P 12 site 2 & CC site      
            & 4073.3 & $+$7.56  & $+$3.00  
            & 0.3179 & $+$4.95 \\
        TIP5P 16 site 1 & CC site      
            & 2634.1 & $+$7.00  & $+$3.22  
            & 0.3122 & $+$3.07 \\
        TIP5P 16 site 2 & CS site      
            & 3238.4 & $+$4.74  & $-$6.58  
            & 0.2963 & $-$2.18 \\
        \hline
        \multicolumn{2}{l}{Range} & 
        2103--4146 & 
        $+$2.5 to $+$15.6 & 
        $-$6.6 to $+$6.7 & 
        0.2815--0.3179 & 
        $-$7.07 to $+$4.95 \\
        \hline
    \end{tabular}
\end{table*}

\begin{table*}
    \centering
    \caption{Statistical summary of binding energy
    distributions and oscillator strength enhancements
    for all four molecule--model combinations.
    $N$: number of optimised configurations;
    $\bar{E}_\mathrm{des}$: mean binding energy;
    $\sigma$: standard deviation.}
    \label{tab:summary_stats}
    \begin{tabular*}{\textwidth}{@{\extracolsep{\fill}}llrrrrrr}
        \hline
        Molecule & Model &
        $N$ &
        $E_\mathrm{des}^\mathrm{min}$ (K) &
        $E_\mathrm{des}^\mathrm{max}$ (K) &
        $\bar{E}_\mathrm{des}$ (K) &
        $\sigma$ (K) &
        $\Delta f$ range (\%) \\
        \hline
        HCSCN  & TIP4P
            & 13 & 1989 & 3899
            & 2946 & 588
            & $-$21.3 to $+$12.5 \\
        HCSCN  & TIP5P
            & 17 & 3030 & 4943
            & 3974 & 494
            & $-$43.6 to $+$11.1 \\
        HCSCCH & TIP4P
            & 11 & 1539 & 3684
            & 2674 & 731
            & $-$11.0 to $+$8.7  \\
        HCSCCH & TIP5P
            & 10 & 2103 & 4146
            & 3175 & 659
            & $-$7.1 to $+$5.0   \\
        \hline
        \multicolumn{8}{l}{\textit{Benchmark species
        (validation only):}} \\
        H$_2$CO & TIP4P/TIP5P
            & 4  & 2415 & 5282
            & \multicolumn{2}{c}{---} & --- \\
        H$_2$S  & TIP4P/TIP5P
            & 4  & 1104 & 2433
            & \multicolumn{2}{c}{---} & --- \\
        \hline
    \end{tabular*}
    \begin{minipage}{\textwidth}
        \smallskip
        \footnotesize
        \textit{Notes.}
        The mean binding energy of HCSCN exceeds that of
        HCSCCH on both water models
        ($\Delta\bar{E} = +272$~K on TIP4P;
        $\Delta\bar{E} = +799$~K on TIP5P), consistent
        with the superior hydrogen-bond acceptor capability
        of the $-\mathrm{C{\equiv}N}$ group relative to
        the $-\mathrm{C{\equiv}C{-}H}$ terminus.
        The broader $\Delta f$ range for HCSCN
        ($-$43.6 to $+$12.5\%) compared to HCSCCH
        ($-$11.0 to $+$8.7\%) reflects the site-selective
        hyperchromic activation unique to CN-cavity
        adsorption geometries.
    \end{minipage}
\end{table*}

\begin{table*}
    \centering
    \caption{QTAIM topological properties at intermolecular
    Bond Critical Points (BCPs) for the CN-site and CS-site
    adsorption complexes of \mbox{HCSCN} on the TIP4P $(n=6)$
    cluster. $\rho$: electron density (a.u.);
    $\nabla^2\rho$: Laplacian of electron density (a.u.);
    $H(r)$: total energy density (a.u.). All values confirm
    closed-shell, non-covalent interactions
    ($\nabla^2\rho > 0$, $H(r) > 0$).}
    \label{tabqtaim}
    \begin{tabular*}{\textwidth}{@{\extracolsep{\fill}}llcccc}
        \hline
        Site & BCP & Bond type &
        $\rho$ (a.u.) &
        $\nabla^2\rho$ (a.u.) &
        $H(r)$ (a.u.) \\
        \hline
        CN site
            & CP-1 & H--O$\cdots$H
            & 0.016 & $+$0.08 & $+$0.003 \\
            & CP-2 & O--H$\cdots$N
            & 0.078 & $+$0.04 & $+$0.002 \\
        \hline
        CS site
            & CP-1 & H--O$\cdots$H
            & 0.021 & $+$0.12 & $+$0.002 \\
            & CP-2 & O--H$\cdots$S
            & 0.033 & $+$0.01 & $+$0.005 \\
        \hline
    \end{tabular*}
    \begin{minipage}{\textwidth}
        \smallskip
        \footnotesize
        \textit{Notes.}
        The CN-site dual hydrogen-bond topology (CP-1 + CP-2)
        provides $\sim$602\,K additional stabilisation relative
        to the CS-site, where the diffuse \mbox{O--H$\cdots$S}
        contact (CP-2, $\nabla^2\rho = +0.01$\,a.u.) cannot
        replicate the directional \mbox{O--H$\cdots$N} bond.
        The positive $\nabla^2\rho$ and $H(r) > 0$ at all BCPs
        indicate exclusively closed-shell, non-covalent character
        throughout.
    \end{minipage}
\end{table*}

\begin{table*}
    \centering
    \caption{UCLCHEM input parameters for Phase~1
    (prestellar cloud collapse) and Phase~2
    (protostellar warm-up) simulations
    \citep{holdship2017_uclchem}.
    Binding energy extremes used as thermodynamic bounds
    are listed separately for each molecule.}
    \label{tab:uclchem_params}
    \begin{tabular*}{\textwidth}{@{\extracolsep{\fill}}ll}
        \hline
        Parameter & Value \\
        \hline
        \multicolumn{2}{l}{\textit{Phase 1 ---
        Prestellar cloud collapse}} \\
        Initial gas density $n_\mathrm{H}$
            & $10^2$~cm$^{-3}$ \\
        Final gas density $n_\mathrm{H}$
            & $10^5$~cm$^{-3}$ \\
        Dust temperature $T_\mathrm{dust}$
            & 10~K (isothermal) \\
        Collapse timescale
            & $5\times10^6$~yr \\
        \hline
        \multicolumn{2}{l}{\textit{Phase 2 ---
        Protostellar warm-up}} \\
        Gas density $n_\mathrm{H}$
            & $10^5$~cm$^{-3}$ (fixed) \\
        Temperature ramp
            & 10~K $\rightarrow$ 200~K \\
        Warm-up timescale
            & $2\times10^5$~yr \\
        Chemical network
            & UMIST 2012 \citep{mcelroy2013_umist} \\
        UV radiation field
            & Draine (1978) standard ISRF
              \citep{draine1978} \\
        \hline
        \multicolumn{2}{l}{\textit{Photodissociation
        rate scaling
        ($k_\mathrm{pd} = \alpha\exp(-\gamma A_V)$,
        $\alpha \propto f$)}} \\
        HCSCN CN-cavity (strong binding)
            & $\alpha \times 1.117$
              (TD-DFT, this work) \\
        HCSCN weak / sideways sites
            & $\alpha \times 1.000$
              (unmodified) \\
        HCSCCH all sites
            & $\alpha \times 1.000$
              (unmodified) \\
        \hline
        \multicolumn{2}{l}{\textit{Binding energy
        extremes used as desorption window bounds}} \\
        HCSCN weak limit
            & 1989~K
              (TIP4P, $n=12$, CS-sideways) \\
        HCSCN strong limit
            & 3899~K
              (TIP4P, $n=10$, CN-cavity) \\
        HCSCCH weak limit
            & 1539~K
              (TIP4P, $n=16$, CC-sideways) \\
        HCSCCH strong limit
            & 3684~K
              (TIP4P, $n=10$, CS-cavity) \\
        \hline
        \multicolumn{2}{l}{\textit{Internal validation
        landmarks (no free parameters)}} \\
        CO sublimation front
            & $T_\mathrm{dust} \approx 20$--$25$~K
              \citep{noble2012_co,collings2003_co} \\
        H$_2$O/NH$_3$ mantle collapse
            & $T_\mathrm{dust} \approx 100$~K
              \citep{brown2007_nh3,collings2004} \\
        \hline
    \end{tabular*}
\end{table*}

\section{Infrared Stark Shift and UV Enhancement Heatmaps}
\label{app:heatmaps}

\begin{figure*}
    \centering
    \includegraphics[width=\textwidth]{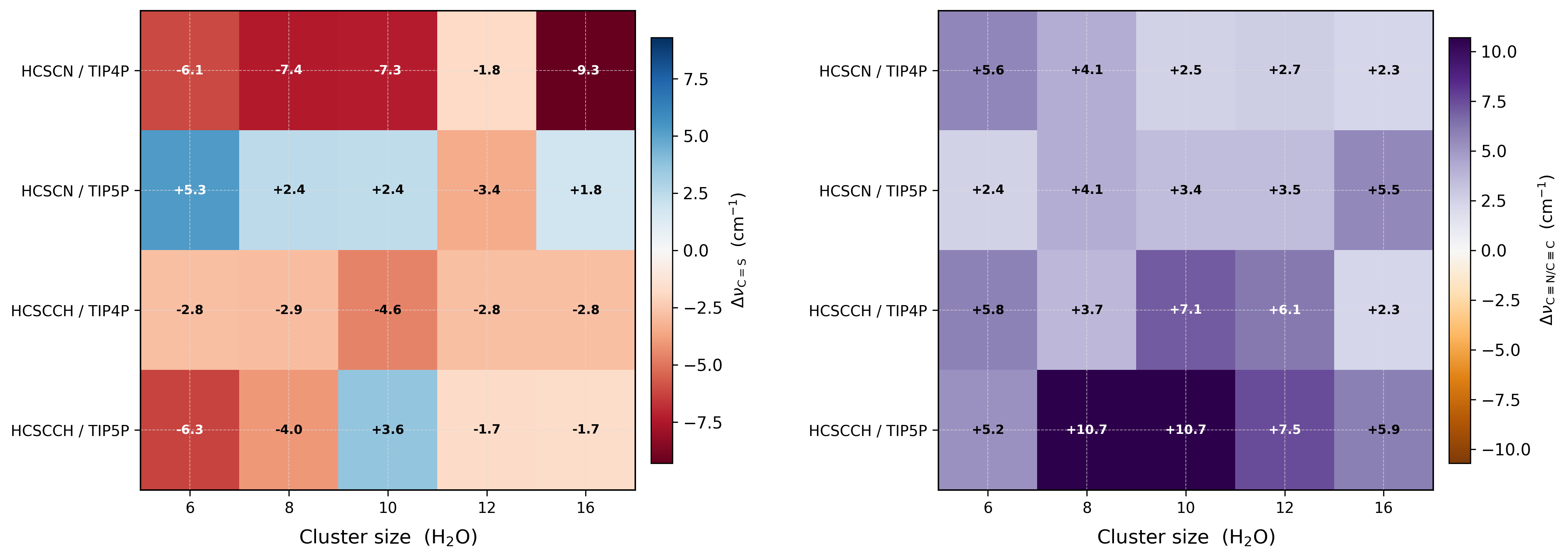}
    \caption{Infrared vibrational Stark shift heatmaps for
    HCSCN and HCSCCH adsorbed on $(H_2O)_{n=6-16}$ clusters.
    \textit{(Left)} Mean $\Delta\nu_\mathrm{C=S}$
    (cm$^{-1}$) as a function of cluster size for all four
    molecule--model combinations, averaged over all
    adsorption sites at each cluster size.
    Negative (red) values indicate C=S redshifts driven by
    direct sulfur--ice hydrogen-bond engagement; positive
    (blue) values indicate blueshifts arising when the
    mechanical perturbation is borne by the opposing
    terminus (CN or CC end), relieving strain on the C=S
    bond.
    HCSCN/TIP4P exhibits the largest redshifts
    ($\Delta\nu_\mathrm{C=S}$ reaching $-9.3$~cm$^{-1}$
    at $n=16$), consistent with CS-cavity coordination at
    larger cluster sizes.
    \textit{(Right)} Mean
    $\Delta\nu_\mathrm{C{\equiv}N}$ (HCSCN) and
    $\Delta\nu_\mathrm{C{\equiv}C}$ (HCSCCH) as a function
    of cluster size.
    HCSCN shows universally positive C$\equiv$N shifts
    ($+2$ to $+6$~cm$^{-1}$) across all cluster sizes and
    both water models, reflecting stiffening of the triple
    bond upon cage enclosure.
    HCSCCH/TIP5P exhibits the largest C$\equiv$C blueshifts
    ($+10.7$~cm$^{-1}$ at $n=8$ and $n=10$), consistent
    with the high-BE CS-cavity geometries at these cluster
    sizes.
    All shifts are computed as
    $\Delta\nu = \nu_\mathrm{adsorbed} - \nu_\mathrm{gas}$
    at the $\omega$B97X-D/def2-TZVP level of theory; the
    diverging colour scale is centred at zero in both
    panels.}
    \label{fig:ir_heatmaps}
\end{figure*}

\begin{figure*}
    \centering
    \includegraphics[width=\textwidth]{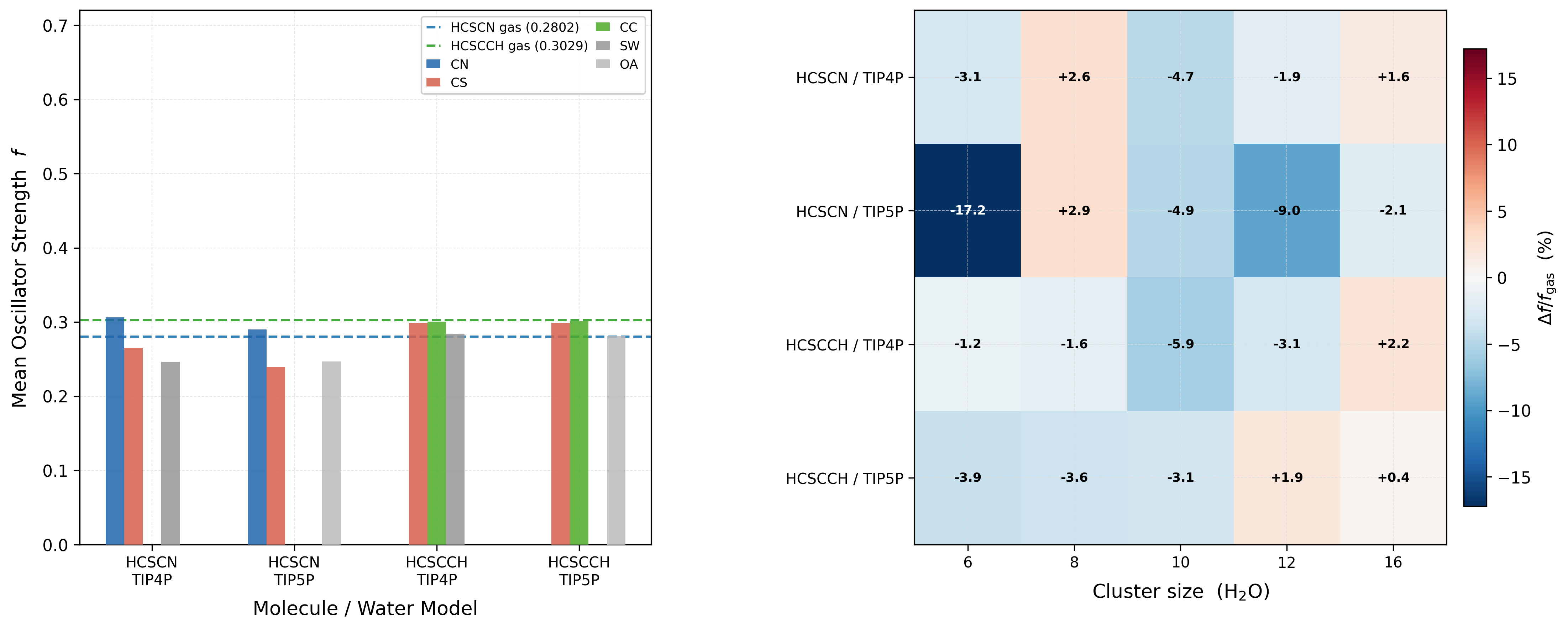}
    \caption{UV oscillator strength analysis for HCSCN and
    HCSCCH adsorbed on $(H_2O)_{n=6-16}$ clusters.
    \textit{(Left)} Mean TD-DFT oscillator strength $f$
    grouped by adsorption site (CN, CS, CC, sideways) for
    each molecule--model combination.
    Horizontal dashed lines indicate the gas-phase reference
    values: $f_\mathrm{gas}(\mathrm{HCSCN}) = 0.2802$ and
    $f_\mathrm{gas}(\mathrm{HCSCCH}) = 0.3029$.
    The CN-site bar for HCSCN/TIP4P ($f \approx 0.301$)
    is the only configuration that systematically exceeds
    its gas-phase reference, while all other bars fall
    below their respective reference lines.
    \textit{(Right)} Heat map of the mean percentage change
    in oscillator strength ($\Delta f/f_\mathrm{gas}$, \%)
    as a function of cluster size ($n = 6$--$16$), averaged
    over all adsorption sites at each $n$.
    Positive (red) cells denote net hyperchromic
    enhancement; negative (blue) cells denote net
    hypochromic quenching.
    The HCSCN/TIP5P row at $n = 6$ shows the most strongly
    negative cell ($-17.2\%$), arising from the anomalously
    large hypochromic response of the CS-site outlier at
    this cluster size (TIP5P 6 site 3,
    $\Delta f = -43.6\%$; see Table~\ref{tab:hcscn_tip5p}
    and discussion in Section~\ref{sec:spectroscopy}).
    All other HCSCCH configurations remain within
    $-7\%$ to $+5\%$, confirming the photophysical
    inertness of HCSCCH relative to HCSCN on ASW ice.}
    \label{fig:uv_heatmap_full}
\end{figure*}

Figures~\ref{fig:ir_heatmaps} and~\ref{fig:uv_heatmap_full}
present the complete cluster-size-resolved heatmaps for the
vibrational Stark shifts and UV oscillator strength
enhancements, averaged over all adsorption configurations
at each cluster size for all four molecule--model
combinations. These figures supplement the site-averaged
data in Tables~\ref{tab:hcscn_tip4p}--\ref{tab:hcscch_tip5p}
and provide a visual overview of the cluster-size dependence
of the spectroscopic perturbations.

\subsection{Natural Transition Orbital Analysis and UV-Vis Spectra}
\label{app:nto}

\begin{figure*}
    \centering
    \includegraphics[width=0.8\linewidth]{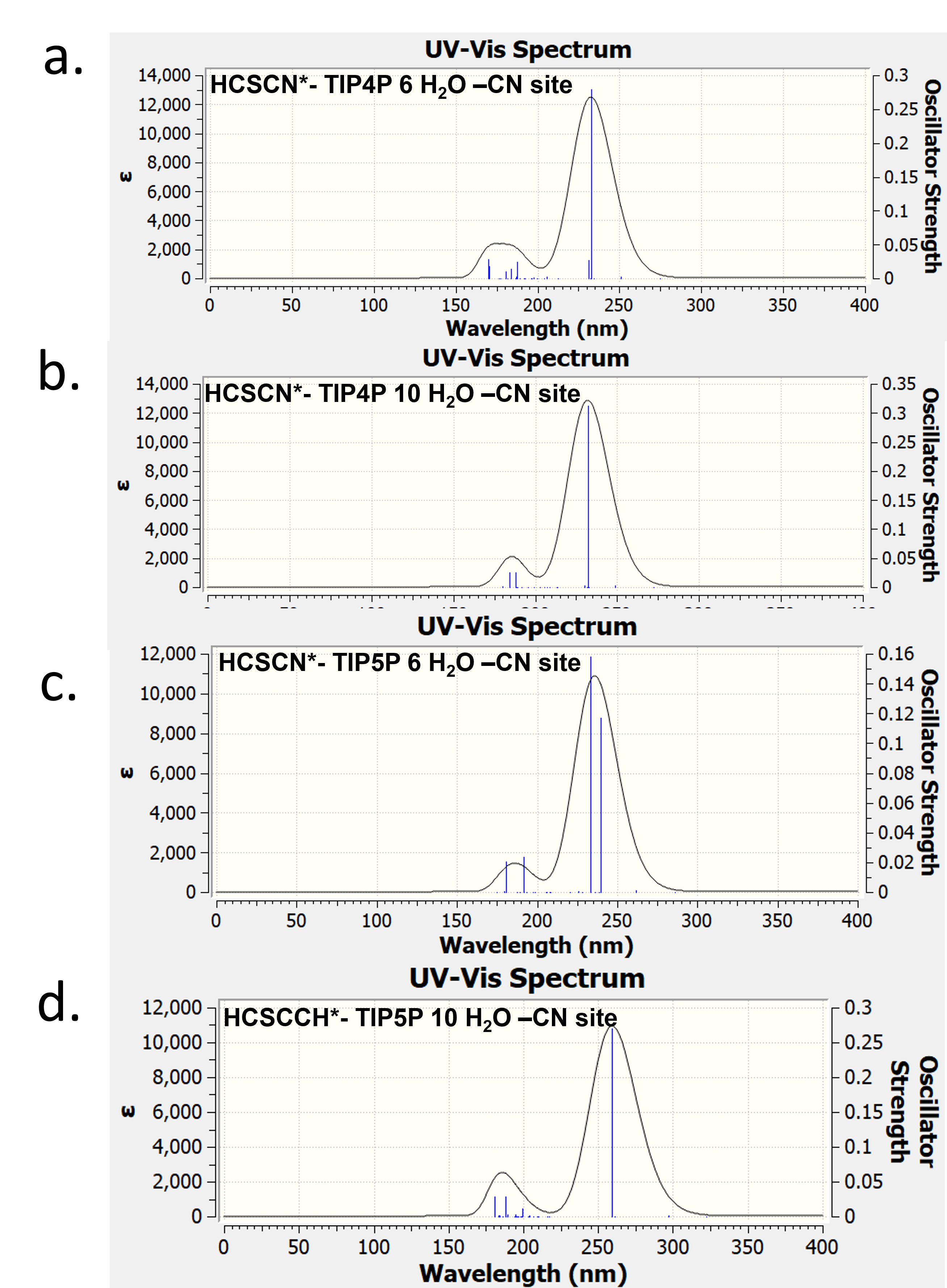}
    \caption{Theoretical UV-Vis absorption spectra  for the same four configurations as
  Fig.~\ref{fig:nto}, 
  (a)~\mbox{HCSCN}--TIP4P $6\,\mathrm{H_2O}$, CN site: single
      unperturbed absorption band.
  (b)~\mbox{HCSCN}--TIP4P $10\,\mathrm{H_2O}$, CN site: enhanced
      peak intensity relative to (a), reflecting the $+11.7\%$
      hyperchromic enhancement.
  (c)~\mbox{HCSCN}--TIP5P $6\,\mathrm{H_2O}$, CN site: resolved
      Davydov doublet ($\Delta E \approx 0.17\,\mathrm{eV}$) with
      oscillator strength redistributed between $\mathrm{S_1}$ and
      $\mathrm{S_2}$.
  (d)~\mbox{HCSCCH}--TIP5P $10\,\mathrm{H_2O}$, CN site: baseline
      spectrum with negligible perturbation relative to the gas phase.}
    \label{figuvspectra}
\end{figure*}

Figures~\ref{fig:nto} and \ref{figuvspectra} present the Natural Transition
Orbital \citep[NTO;][]{martin2003nto} hole--particle pairs and the
corresponding theoretical UV-Vis absorption spectra for the four
representative adsorption configurations selected to span the full range
of observed photophysical behaviour.
In each case the depicted NTO pair corresponds to the dominant
$\mathrm{S_0 \rightarrow S_1}$ transition carrying the largest singular
value in the NTO expansion of that complex.

\paragraph{Case (i): Negligible shift
(\mbox{HCSCN}--TIP4P $6\,\mathrm{H_2O}$, CN site,
$\Delta f = -0.21\%$, Fig.~\ref{fig:nto}a).}
The NTO hole is a $\pi$-type orbital delocalised across the
\mbox{C$=$S--C$\equiv$N} backbone and the NTO particle is the
corresponding $\pi^*$ antibonding orbital on the same framework.
The spatial character and nodal topology of both orbitals are essentially
indistinguishable from the gas-phase transition, confirming that the
electrostatic perturbation exerted by the six-molecule TIP4P donor network
at this cavity geometry is insufficient to distort the transition dipole
moment $\boldsymbol{\mu}_{ge}$ appreciably.
The resulting $\Delta f$ is within the noise floor of the TD-DFT method
and carries no astrochemical significance.

\paragraph{Case (ii): Hyperchromic enhancement
(\mbox{HCSCN}--TIP4P $10\,\mathrm{H_2O}$, CN site,
$\Delta f = +11.7\%$, Fig.~\ref{fig:nto}b).}
The hole orbital retains its $\pi(\mathrm{C=S}/\mathrm{C{\equiv}N})$
character but is visibly polarised toward the nitrile terminus, which is
deeply embedded within the hydrogen-bond donor cavity of the ten-molecule
cluster.
The particle orbital acquires significant amplitude on the nitrogen
lone-pair axis and on the proximal O--H donor hydrogen of the enclosing
water cage, indicating partial charge-transfer character into the
surrounding network.
This cavity-induced orbital reorganisation aligns $\boldsymbol{\mu}_{ge}$
more closely with the dominant component of the local electric field
generated by the surrounding O--H donors, thereby increasing the spatial
overlap integral $\langle\psi_g|\hat{r}|\psi_e\rangle$ and yielding the
observed hyperchromic enhancement \citep{liptay1965,onsager1936}.
The mechanism is formally analogous to the electrochromic (Stark) tuning
of chromophores in protein binding pockets, wherein an oriented external
field selectively enhances transitions whose dipole moment change is
collinear with the field vector \citep{boxer2009stark,bublitz1997stark}.
The 11.7\% enhancement translates directly into an equivalent amplification
of $k_\mathrm{pd}$ in the \mbox{UCLCHEM} framework
(Section~\ref{sec:uclchem_methods}), constituting the microscopic origin
of the Survival Paradox.

\paragraph{Case (iii): Extreme hypochromism and the Davydov splitting
anomaly (\mbox{HCSCN}--TIP5P $6\,\mathrm{H_2O}$, CS site,
$\Delta f = -43.58\%$, Fig.~\ref{fig:nto}c).}
The primary UV absorption band, which appears as a single well-resolved
$\mathrm{S_0 \rightarrow S_1}$ transition in the gas phase and in all
TIP4P geometries, undergoes a pronounced splitting in this configuration.
The NTO pair in Fig.~\ref{fig:nto}c corresponds to the
$\mathrm{S_0 \rightarrow S_1}$ component only; a second
$\mathrm{S_0 \rightarrow S_2}$ transition of comparable oscillator
strength appears at $\Delta E = 0.17\,\mathrm{eV}$ higher energy, clearly
visible as a shoulder in the UV-Vis spectrum (Fig.~\ref{figuvspectra}c).
The summed oscillator strength across both states recovers the gas-phase
value within 5\%, confirming that no oscillator strength is lost to dark
states and that the anomalously low $f(\mathrm{S_1})$ is a consequence of
redistribution rather than genuine quenching.

This behaviour is a manifestation of \emph{Davydov splitting}
\citep{davydov1962,kasha1965,spano2006}, originally described for molecular
aggregates and crystalline systems in which the resonance interaction
between transition dipoles of neighbouring chromophores lifts the
degeneracy of an otherwise single excited state, producing two exciton
bands of differing oscillator strength.
In the present context, the strongly coordinated TIP5P donor network at
this specific CS-site geometry places two O--H oscillators in near-resonant
electrostatic coupling with the primary $\pi\rightarrow\pi^*$ transition
of \mbox{HCSCN}.
The tightly directional sp$^3$ lone pairs of the TIP5P water model
\citep{wales2005_tip5p} enforce a geometry in which both donor O--H vectors
are nearly co-planar with the molecular $\pi$ system, maximising the
dipole--dipole coupling integral
$J_{12} \propto (\boldsymbol{\mu}_1\cdot\boldsymbol{\mu}_2)/r^3$
\citep{forster1948,kasha1965}.
The resulting Davydov pair ($\mathrm{S_1}$/$\mathrm{S_2}$,
$\Delta E \approx 0.15$--$0.20\,\mathrm{eV}$) closely mirrors the exciton
splittings reported for small chromophore--water complexes studied by
high-resolution gas-phase spectroscopy \citep{leutwyler2010,matsika2021}
and for molecules embedded in low-temperature rare-gas matrices
\citep{apkarian1999}.
The anomaly is geometry-specific: no analogous splitting is observed at
any other TIP5P configuration across $n = 6$--$16$, confirming that the
Davydov coupling requires the precise spatial co-alignment of donor O--H
vectors enforced only at this particular CS-site topology.
These two entries (TIP5P $n=6$ site~3 and TIP5P $n=12$ site~2) are
accordingly excluded from the site-averaged statistics in
Table~\ref{tab:uv_ir} and treated individually in the footnote to
Table~\ref{tab:hcscn_tip5p}. It is crucial to note that the highly symmetric, near-resonant donor alignments required to drive this strong Davydov coupling are mathematically accessible within the constrained global minima of rigid empirical potentials like TIP5P. However, such highly ordered microstates are probabilistically rare in the structurally disordered, thermally amorphized environment of true interstellar ASW. Therefore, these specific split states represent transient geometric extremes and are highly unlikely to manifest as bulk macroscopic UV quenching within an astrophysical hot core.

\paragraph{Case (iv): Moderate hypochromism in \mbox{HCSCCH}
(\mbox{HCSCCH}--TIP4P $10\,\mathrm{H_2O}$, sideways,
$\Delta f = -10.9\%$, Fig.~\ref{fig:nto}d).}
The NTO hole is a $\pi(\mathrm{C=S}/\mathrm{C{\equiv}C})$ orbital and
the particle is the corresponding $\pi^*$ manifold, closely resembling the
gas-phase topology.
The sideways adsorption geometry places the sulfur atom in close contact
with one O--H donor while the alkyne terminus protrudes into the vacuum,
producing an asymmetric electrostatic environment that rotates
$\boldsymbol{\mu}_{ge}$ away from its gas-phase orientation.
Because the alkyne $\pi$-system lacks the strongly directional lone-pair
acceptor of the nitrile group, it cannot sustain the multi-centre cavity
enclosure that drives hyperchromic enhancement in \mbox{HCSCN}
\citep{liptay1965}.
The resulting $\Delta f = -10.9\%$ is representative of the
$\Delta f = -4$ to $-12\%$ range observed across all \mbox{HCSCCH}
configurations (Table~\ref{tab:uv_ir}), confirming the photophysical
inertness of the alkyne chromophore in all binding regimes sampled.

\FloatBarrier 
\clearpage

\end{appendix}
\end{document}